\documentclass[namedreferences]{solarphysics}

\usepackage[hyperref,optionalrh,natbib]{spr-sola-addons} 
\usepackage{longtable}
\usepackage{graphicx}        
\usepackage{color}           

\newcommand{\aap}{    {\it Astron. Astrophys.}}

\newcommand{\apj}{    {\it Astrophys. J.}}

\newcommand{\mnras}{  {\it Mon. Not. Roy. Astron. Soc.}}

\newcommand{\solphys}{{\it Solar Phys.}}

\begin{document}

\begin{article}

\begin{opening}

\title{Angular dependence of the facular-sunspot coverage relation as derived by  MDI magnetograms}
\author{S.~\surname{Criscuoli}$^{1}$     
       }
\runningauthor{Criscuoli}
\runningtitle{Angular dependence of facular-sunspot relation}

   \institute{$^{1}$ National Solar Observatory - 3665 Discovery Dr., Boulder, CO 80303, USA\\
                     email: \url{scriscuo@nso.edu} \\ 
                        }

\begin{abstract}
Previous studies have shown that the variation over the solar magnetic activity cycle of the area of facular/network features 
identified on broad band and narrow band imagery is positively correlated with 
the sunspot area and number, the relation between the area coverages being described as either linear or quadratic.
On the other hand, the temporal variation of the spatial distributions of faculae, network and sunspots \textbf{follows patterns} that are less obviously 
correlated, so that we expect the relation that describes variation of the area coverage of different types of magnetic features to vary with
the position over the disk. 
In this work we employ MDI full-disk magnetograms acquired during Cycle 23 and at the beginning of Cycle 24 to investigate 
the relation between the coverage of magnetic elements characterized by different \textbf{amounts} of magnetic flux and located at different angular distances from disk center  
with the sunspot number. In agreement with some previous studies we find that daily data are best described by a quadratic function while data averaged over 6-months
are best described by a linear function. In both cases the coefficients of the fits show large dependence on the position over the disk 
and the magnetic flux. We also find that toward disk center 6-months averaged data show asymmetries between the ascending and the descending phase. 
Implications for solar irradiance modeling are discussed. 

\end{abstract}
\end{opening}

\section{Introduction}
\label{intro}
Sunspot, faculae and network are manifestations of the solar magnetic field.
Since earlier observations it is known that the number and position of sunspots
varies with a periodicity of approximately 11 years \citep{schwabe1844,carrington1858,maunder1904}. 
Subsequent studies have shown that the area coverage of 
faculae and network is in general positively correlated with the magnetic activity 
\citep[see][for a recent review]{ermolli2014}
although in the literature there is some discrepancy in the description of this correlation, mostly because of the different datasets and data processing
employed by different authors \citep[e.g.][]{foukal1996,fliggesolanki1998,ermolli2009,bertello2010,gyori2012}.   Several studies showed that the relation
between area of
facular/network regions identified on chromospheric and photospheric imagery and sunspot area can be described with a linear function
 \citep[e.g.][]{brownevans1980,foukal1993,chapman1997,chapman2011,priyal2014}, with the coefficients of the fit being dependent on the strength of the cycle\citep{brownevans1980}, while some 
authors described the relation as quadratic \citep{foukal1993, fliggesolanki1998}.  \citet{brownevans1980} analyzed the variation of white light faculae  
using data acquired over nine cycles at the Royal Greenwich Observatory and noticed that the relation between facular/network and sunspot area is linear during
 the ascending and descending phases of the cycle, while during the maximum it is difficult to find a clear relation, so that the relation seems to be best described with a
 sigmoid. Those authors also noticed that in certain cycles 
 the area coverage of facular/network regions differs during the ascending and descending phases.

 The spatial distribution of magnetic activity is also known to vary in time \citep[see][for a recent review]{hathaway2010}. Magnetic regions emerge as \textbf{bipoles} at the
 beginning of the cycle preferentially at middle latitudes ($\pm$ 30 deg, activity belt)
 and then closer to the equator as the cycle proceeds, while advective processes transport the flux of the following polarity preferentially toward the poles 
 \citep[e.g.][for a recent review]{mackay-yeates2012} mainly in the form of faculae and network.  
 As a result, the variation of facular and network coverage over the cycle is also strongly dependent on their latitudinal distribution and the associated magnetic
 flux intensity \cite[e.g.][]{mullerroudier1984,hagenaar2003,sheeley2011,jin2011,jinwang2012,petrie2014}.
 Polar faculae are for instance
 anti-correlated with the sunspot number \citep[e.g.][for a recent review]{petrie2014}. The area of faculae is positively correlated with sunspot area and number and 
 their coverage closely follows the latitudinal migration of sunspots. For what concerns the network, results are instead still controversial.
   Recently, \citet{jin2011} and \citet{jinwang2012} found that, depending on the amount of the associated magnetic flux, the number of network elements can be 
  correlated, anti-correlated, or constant over the activity cycle, with the component positively correlated also being preferentially distributed over the activity belt. 
  Some analysis carried out on Ca II K images indicate that the area coverage of the 
 so called ``quiet'' and ``active'' network
 are anti-correlated and correlated, respectively, with the magnetic activity cycle \citep[e.g.][]{mullerroudier1984,ermolli2011, fontenla2011}, while some others indicate
 that the network coverage is always positively correlated \citep[e.g.][]{worden1998,singh2012,priyal2014}. 
 A recent analysis carried out by \citet{utz2016} on sub-arcsec spatial resolution G-band data, also indicates that the variation of G-band magnetic bright points observed at disk center
 is positively correlated with the 11-year cycle.

Studies of the variation of facular, network and sunspot coverage over the magnetic cycle are of relevance for solar irradiance studies, as several techniques
developed to reproduce measured total and spectral solar irradiation variations make use of observations of variations of the solar surface magnetism 
\citep{domingo2009,ermolli2013,yeo2014}. Records of sunspot number are available since the beginning of the 17th century, while records of facular and network 
coverages date back to the beginning of the 20th century. For this reason,  
some techniques developed to infer long-term (decades, centuries) solar irradiance
variations make use of the empirical relation between sunspot and facular/network coverage derived from available data 
\citep[e.g.][]{fligge2000,solanki2002,krivova2007,krivova2011}. The ratio between facular and sunspot area is also a parameter employed in models aimed to reproduce the
photometric variability of stars \citep[e.g.][]{lanza2006, bonomo2008, dumusque2015}.  

Both observational \citep[e.g.][]{ortiz2002,ermolli2007,ermolli2010,yeo2013}
and theoretical \citep[e.g.][]{spruit1976, steiner2005, criscuoli2009,criscuoli2014} studies indicate that the radiative 
emission of magnetic elements is strongly dependent on their angular distance from the center of the solar disk. Moreover, different types 
of magnetic features, in irradiance studies usually classified according to their magnetic flux \citep[e.g.][]{krivova2007,ball2012,yeo2013} or their emission in Ca II K 
\citep[e.g.][]{ermolli2007,ermolli2010,fontenla2011}, are characterized by different photometric properties. 

For solar variability studies it is therefore important to investigate not only the variation of the coverage of faculae and network during the cycle,
as has been 
done in previous studies, but
also to discriminate between different types of features and their location over the disk. 
In this work we investigate the variation over Cycle 23 of the area coverage of magnetic elements characterized by different amount of magnetic flux and located
at different distances from disk center. The paper 
is organized as follows:
in Section~\ref{Sec.Obs} we describe the dataset analyzed; in Section~\ref{Sec.Res} we present our results and in Section~\ref{Sec.Noi} we investigate the effects of noise;
in Section~\ref{Sec.Concl} we discuss our results in the light of previous works and we draw our conclusions.

\section{Observations} \label{Sec.Obs}

We analyzed a set of 3265 full-disk line-of-sight magnetograms acquired daily between 1996 June 1 and 2011 April 4 with the Michelson Doppler Interferometer (MDI) 
on board the SOHO space craft \citep{scherrer1995}.
The MDI acquires filtergrams in two orthogonally polarized states at five wavelength positions sampling the Ni I 676.8 nm line.
The MDI data-products (intensitygrams, line-depth, line-of-sight velocities and line-of-sight magnetograms) are obtained by a combination of these filtergrams
under the assumption that the shape of the Ni I line can be approximated by a Gaussian 
function \citep{scherrer1995}.
This assumption causes uncertainties in the estimate of the data-products, especially in presence of large magnetic fields, which, 
via the Zeeman effect, cause broadening and saturation of the line \citep[e.g.][]{wachter2006,rajaguru2007,mathew2007,criscuoli2011,couvidat2012}.
For this reason, several efforts have been dedicated to the inter-calibration of MDI line-of-sight magnetograms with data acquired from different instruments 
\citep[e.g.][]{tran2005,demidov2009,ulrich2009,liu2012,pietarila2013}.
In this study we employ five-minutes
level 1.8.2 data, obtained applying the correction factors reported in \citet{ulrich2009}, as described in \citet{liu2012}.

For our analysis we selected magnetic pixels that do not belong
to sunspots and where the absolute value of the magnetic flux $|B|$ exceeds the noise threshold, $Th$. This last quantity was defined as $3\cdot\sigma$,
where $\sigma$ is the map 
of the noise level over the detector.
The map $\sigma$  was estimated
following a procedure similar to the one described in \citet{ortiz2002}. We considered a set of 50 magnetograms
acquired during very quiet days in the period 1996-1997 and 2006-2011. For each magnetogram pixel we computed the standard deviation of magnetic flux values
over a running box of $100\times 100$ pixels;
for pixels closer to the limb the box was reduced to $50\times50$ pixels. Note that, to exclude residual activity,  the standard deviation over each box was computed using the IDL
procedure robust\_sigma.pro, which employs a ``robust Tukey's biweight'' algorithm \citep[e.g.][]{press1992} to discard outliers values.     
The map $\sigma$ was then computed as the average of the maps derived from each magnetogram.
 \citet{ball2012} showed that the noise level of MDI magnetograms 
slightly increases with time,
most likely because of uncertainties in the calibrations after the two periods of loss of contact of the SOHO spacecraft in 1998 and 1999. For this reason we utilized 
two different noise maps for data acquired before and after 1999, the first computed from quiet days acquired in the period 1996-1997 and the second one computed from
quiet days acquired between 2006 and 2011. In agreement with previous studies, we found that the noise level so defined increases toward the limb, decreases drastically toward the extreme limb,
and that a quadrant of the detector shows a systematically 
higher noise level \citep[e.g.][] {ortiz2002, ball2012, liu2012,pietarila2013}. The value of the noise level derived from the 1996-1997 data varied between 6G and 12G,  
while the noise level derived from the 2006-2011 data varied between 9G and 18G. 
The effects of the noise on our results are investigated in Section~\ref{Sec.Noi}.

\begin{figure}
\begin{centering}
\centerline{\includegraphics[width=6.3cm,trim=0mm 0 50mm 0mm, clip=true]{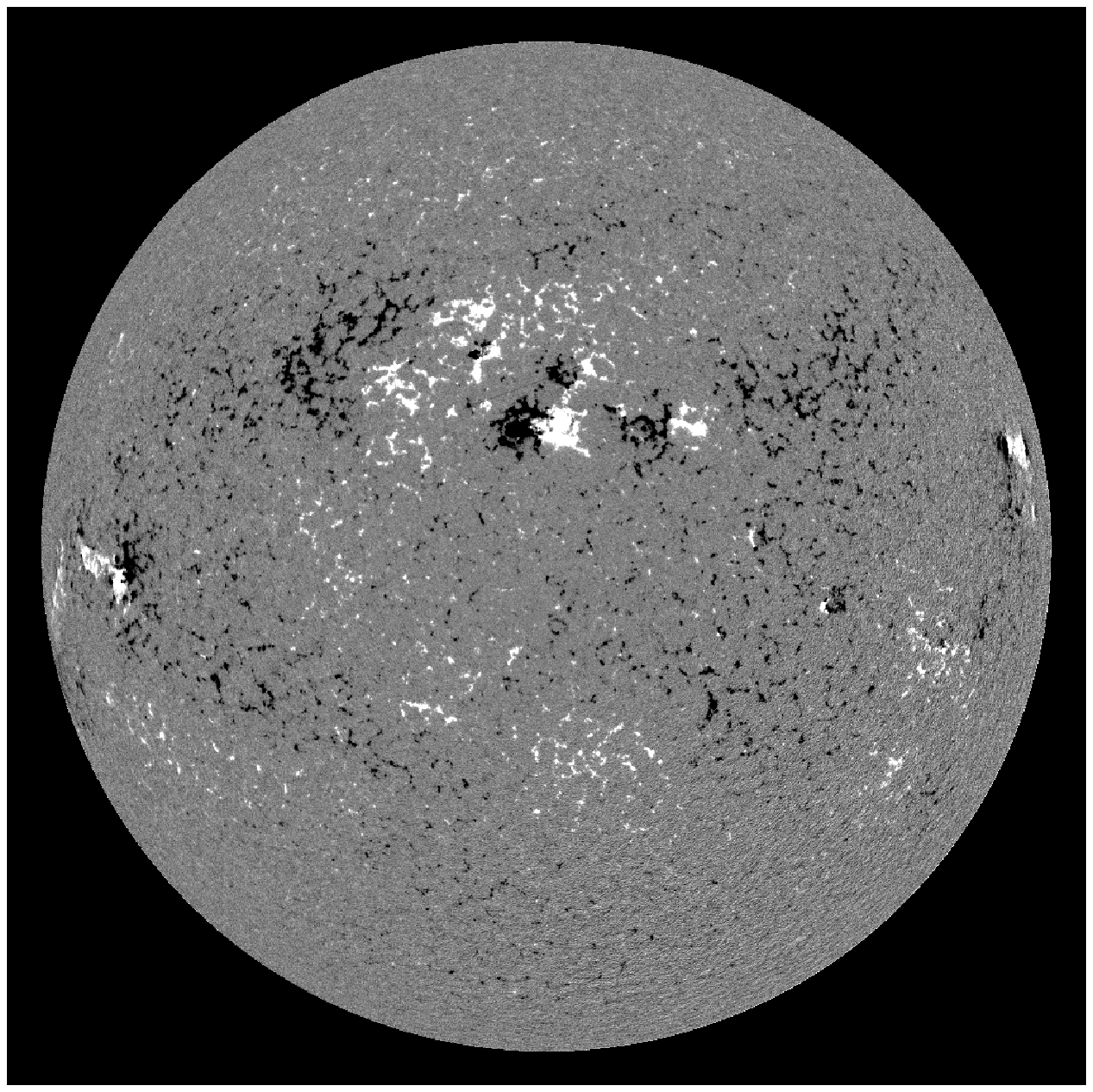}
\includegraphics[width=7.5cm,trim=0mm 0 25mm 0mm, clip=true]{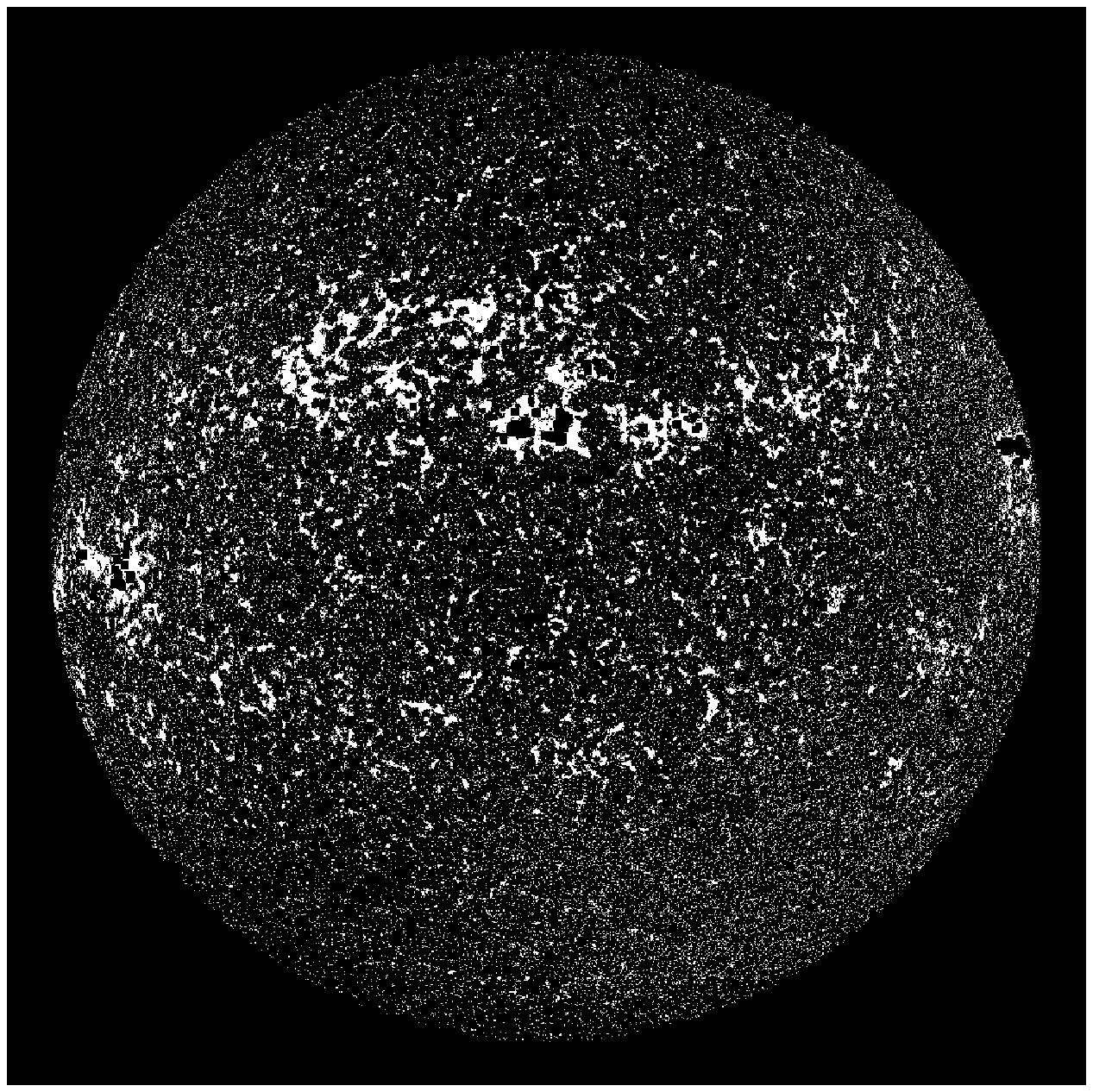}}
\caption{Magnetogram (left) saturated at $\pm$100G and the corresponding mask (right) acquired on July 11 2003. }
\label{fig1}
\end{centering}
\end{figure}

The vertical component of the magnetic flux of the pixels exceeding the noise threshold was estimated as the ratio between the 
measured magnetic flux  and the cosine of the heliocentric angle $\mu$. Sunspot umbrae and penumbrae were then identified by selecting pixels where the
magnetic flux $|B|/\mu$ is larger than 850 G. Isolated pixels were removed applying an ``opening'' operation with a 2-pixels kernel, then the mask was  
``dilated'' with a 8-pixels kernel to improve the selection of penumbral pixels. 
Figure~\ref{fig1}  shows an example of the magnetic pixels selected using the described procedure, together with the original magnetogram. As in previous studies,
the selected features include both network and facular regions. Note that regions where $\mu < 0.2$ were excluded from the analysis.

\begin{figure}
\begin{centering}
\includegraphics[scale=.65]{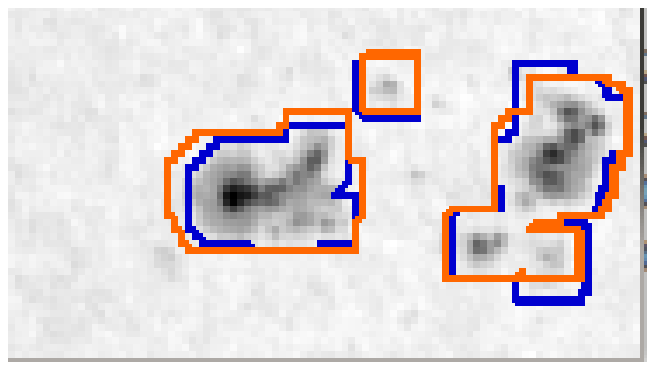}
\includegraphics[scale=.65]{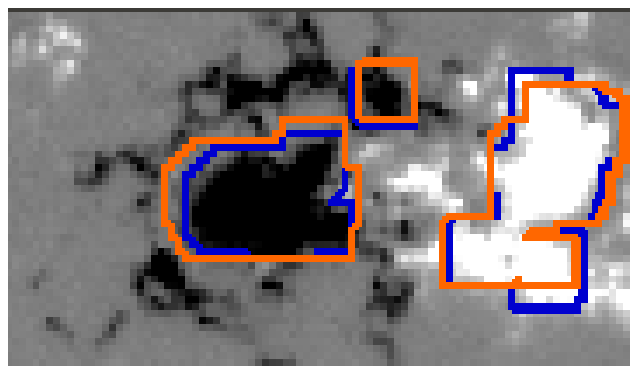}
\caption{Example of the contours of the sunspot masks derived from magnetograms (blue) and intensitygrams (orange) overlaid to a detail of the intensitygram (left)
and magnetogram (right) acquired on July 11 2003. The magnetogram was saturated at $\pm 500G$.}
 \label{fig1a}
\end{centering}
\end{figure}

In some previous studies that investigated properties of faculae and network 
regions through the analysis of MDI magnetograms, umbral and penumbral pixels identified on intensitygrams acquired by the MDI and rotated to match
the time of observations of the magnetograms. 
Nevertheless, we found that several of the available data were corrupted so that for most of the days the 
intensitygrams were acquired  several hours apart from the magnetograms. 
To avoid uncertainties due to interpolations when rotating the intensitygrams we therefore decided to analyze magnetograms only. 
To test the differences between masks obtained with intensitygrams and magnetograms, we compared masks obtained on intensitygrams during days in which
intensitygrams and magnetograms acquired less than one hour apart were available.
The mask from the intensitygrams  was created by first computing contrast images obtained compensating the intensitygrams for the intensity center to limb 
variation of quiet pixels, then pixels where the contrast is lower than 0.85  were selected. The resulting mask was then ``opened'' with a 2-pixels kernel, 
and ``dilated'' with a 8-pixels kernel.
We found that the masks produced in these two ways are overall similar, as shown by comparing the masks contours lines illustrated in Figure\ref{fig1a}.  

Finally, as a proxy of the magnetic activity we employed the revised sunspot number catalog \citep{clette2014} from the
World Data Center SILSO, Royal Observatory of Belgium, Brussels \citep{silso}.

\section{Results} \label{Sec.Res}
For each day we computed the area coverage (in millionth of the visible solar hemisphere, $\mu Hem$) 
over eight annuli centered at disk center, equally spatied in $\mu$, of features characterized by different amount of magnetic flux. Note that, 
unlike in some previous studies, we did not compensate the areas for line-of-sight effects. The analysis was restricted to features with magnetic flux in the range 
$100G<|B|/\mu \le 800G$. The upper limit value was set by the threshold value employed to define sunspot regions. 
The value of the lower limit was defined as the minimum value of the magnetic field that exceeded the $3\cdot \sigma$ at $\mu=0.2$. Being the minimum value 
of $\sigma$ $\simeq$ 6G,  we found a lowest magnetic field value of $\simeq$90G. We employed such definition to reduce statistical inhomogeneity at different
angular distances that derive from the magnetic flux compensation of line-of-sight effects. Note that this approach does not reduce effects due to the non-uniformity of 
the noise over the detector, as discussed in Section~\ref{Sec.Noi}. 

\begin{figure}
\begin{centering}
   \centerline{
               \includegraphics[width=0.48\textwidth,clip=]{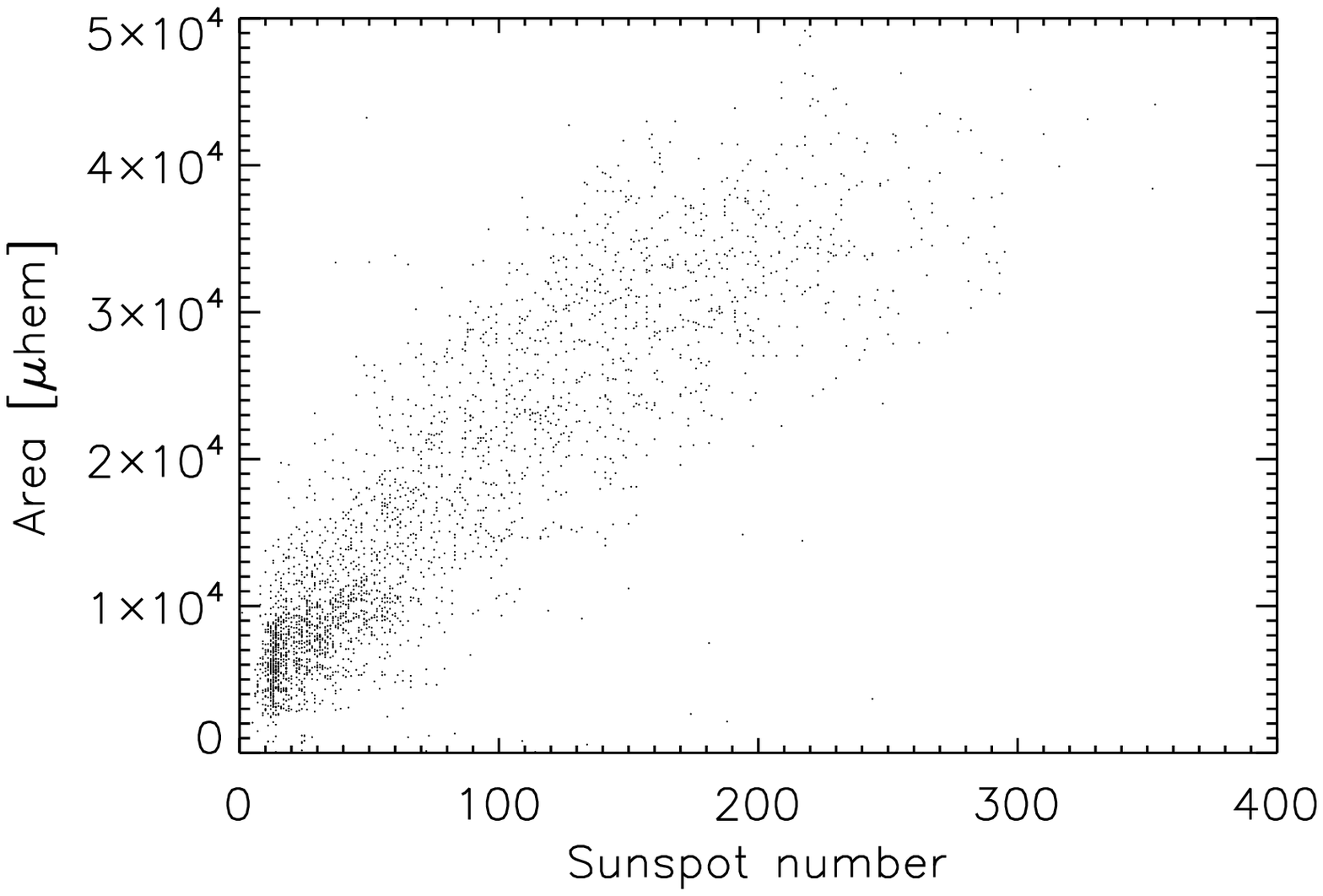}  \includegraphics[width=0.48\textwidth,clip=]{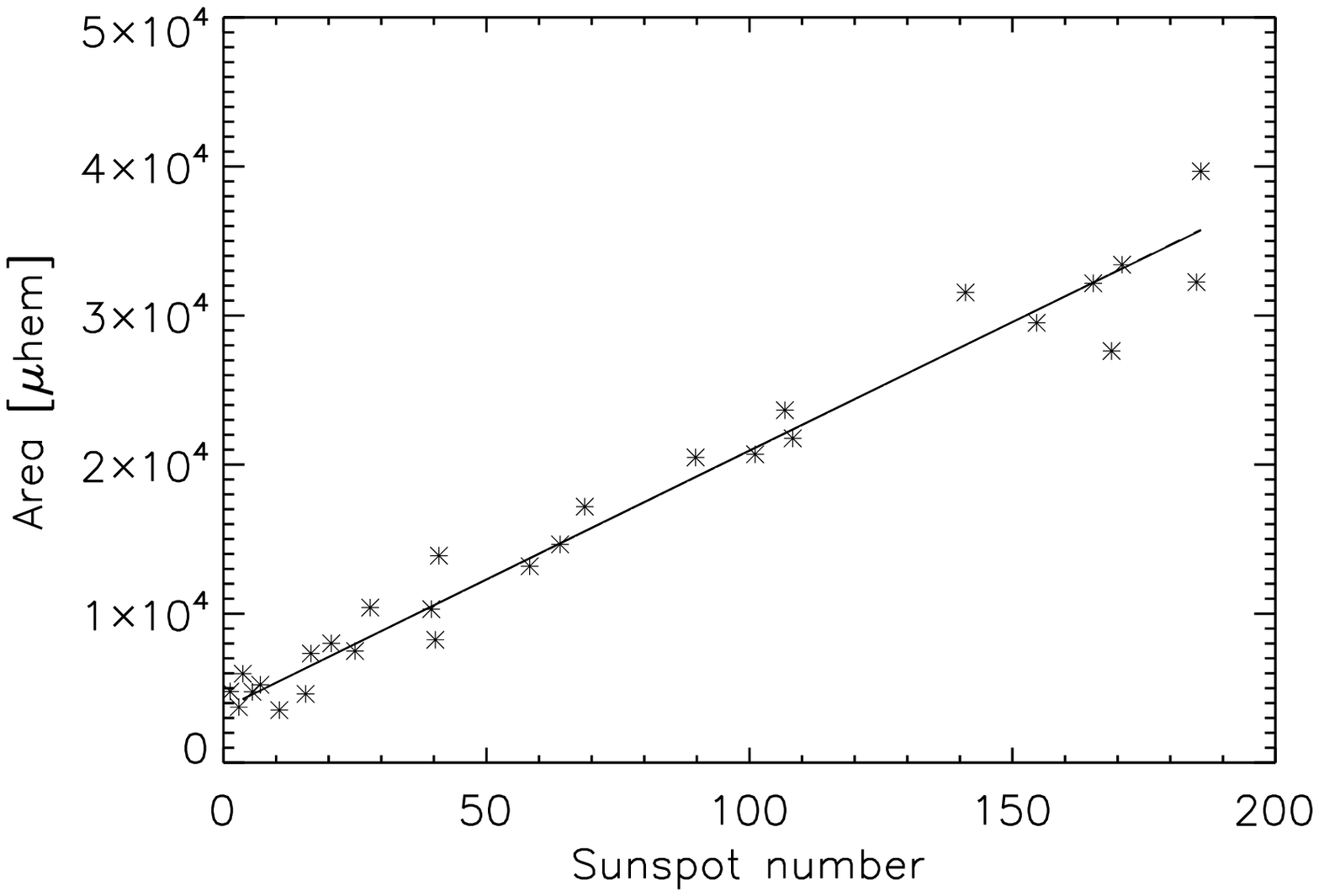}}

     \centerline{  \includegraphics[width=0.48\textwidth,clip=]{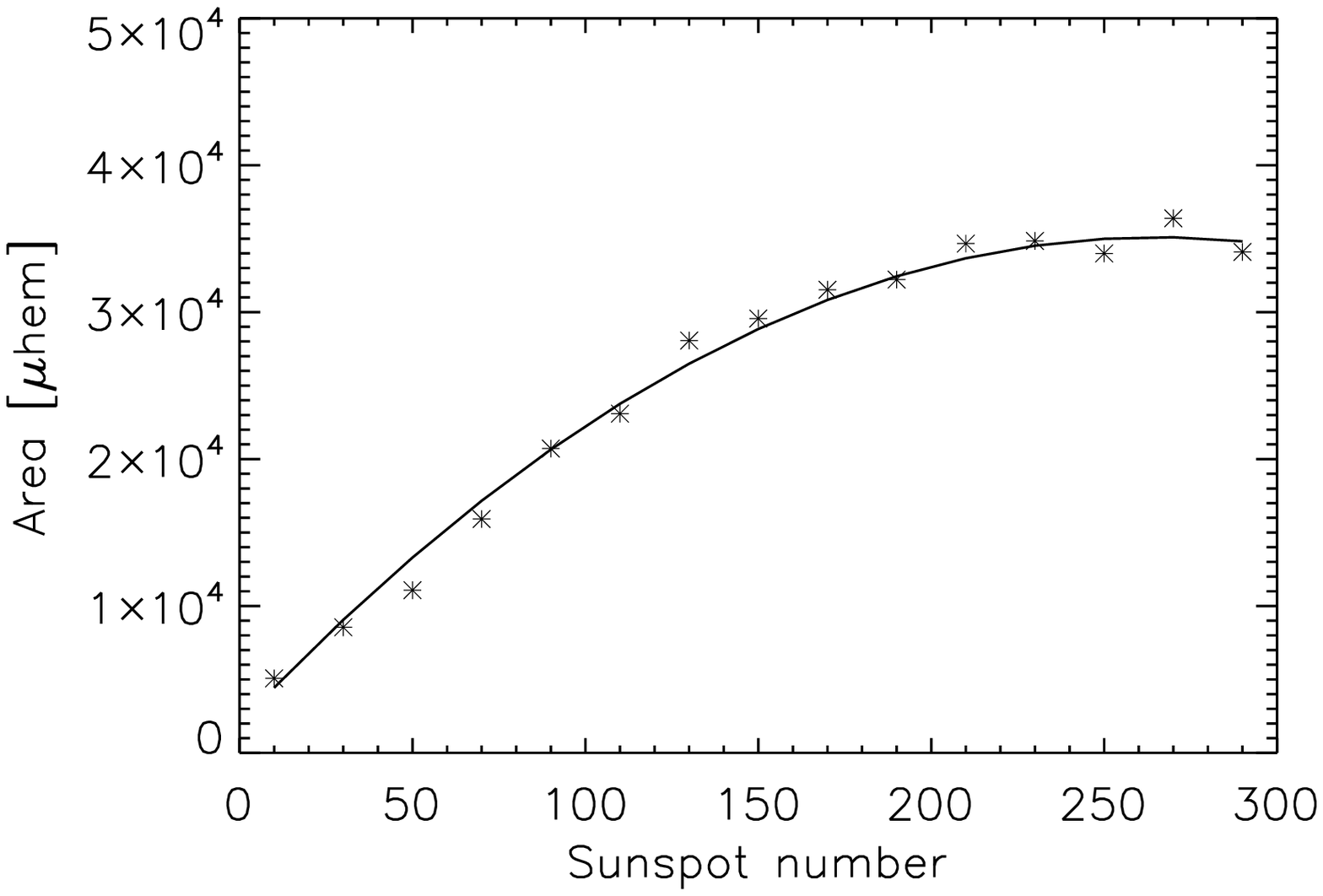} \includegraphics[width=0.48\textwidth,clip=]{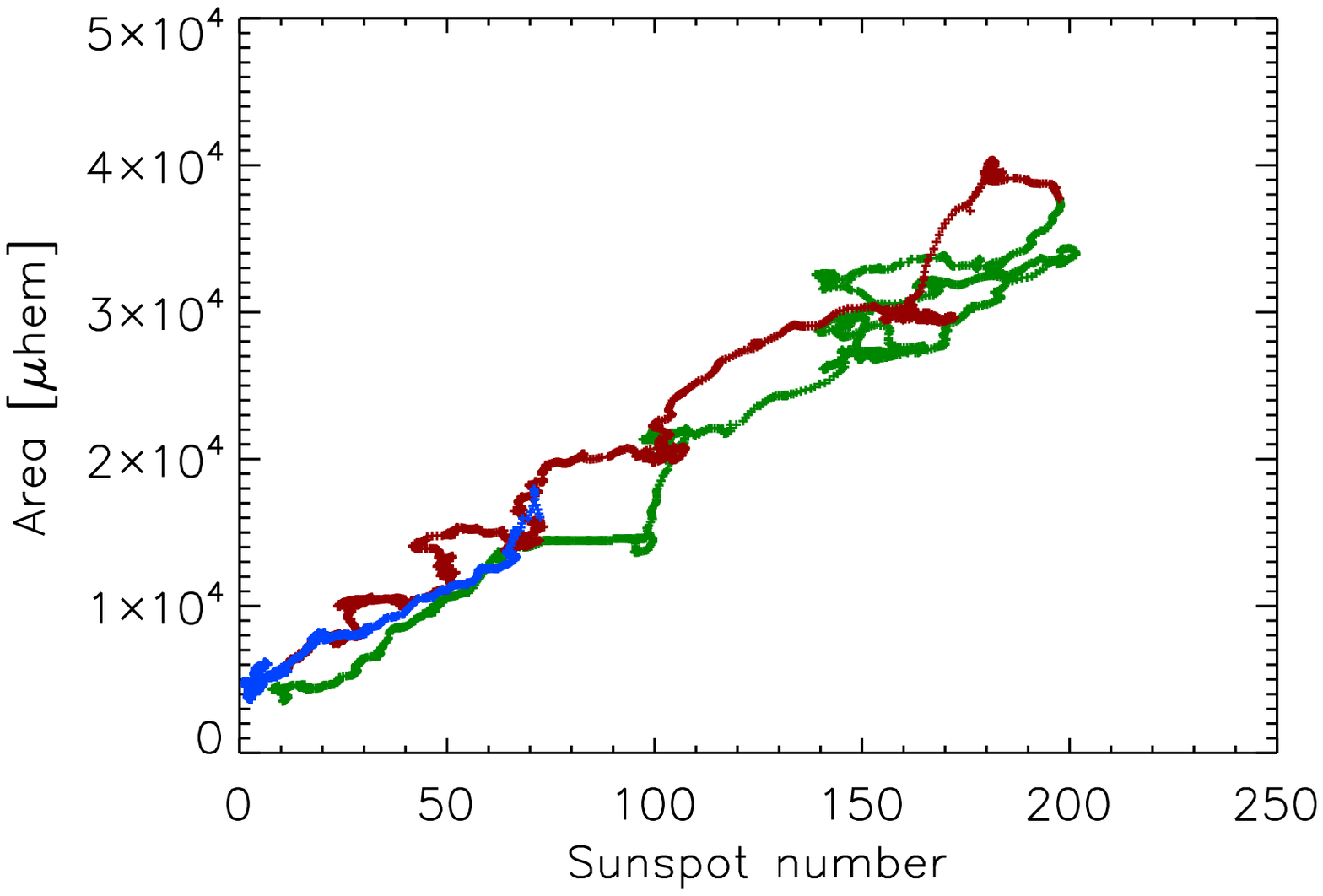}}

\caption{Magnetic pixels area versus sunspot number for pixels located between $0.2<\mu\le1$ and $100G < |B|/\mu \le 800G$. Top-left: daily data. Top-right: 6-months
average data.
Bottom-left: data averaged over 20-sunspot number bins. Bottom-right:data averaged over a six months running window.  Green: data from
May 1996 to June 2002. Red: data from June 2006 to June 2009. Blue: data from June 2009 to April 2011. Lines represent linear (continuous) and a quadratic (dashed)
fits.  }
\label{fig2}
\end{centering}
\end{figure}

\begin{table}[h]
 \caption{Coefficients values returned by the linear and the quadratic fit, together with the Spearman's coefficient R and the F-test value, of data reported in Figure~\ref{fig2}.}
 \label{tbl-maincoeffs}
 \begin{tabular}{c|ccc|cc}
  \hline
  & $\alpha$ & $\beta$ & $\gamma$ & R & F-test \\
  \hline
  DAILY & 6239$\pm$1900 & 116$\pm$11 & - & & \\
         & 1987$\pm$2400 & 250$\pm$50 & -0.47$\pm$0.16& 0.96& 178\\
         \hline
  6-MONTHS AVG. & 3635$\pm$643 & 173$\pm$12 & - & \\
                & 3618$\pm$880 & 174$\pm$44 & -0.007$\pm$0.26 & 0.99 & 0.26\\
                \hline
 \end{tabular}

\end{table}

Figure~\ref{fig2} shows the variation with the sunspot number of the area coverage of pixels
with $100G<|B|/\mu \le 800G$ and located at $0.2 < \mu \le 1$. In particular, the top-left panel shows results obtained from the daily observations, the top-right
panel shows results averaged
over 6 months and the bottom-left panel shows daily data averaged over bins of 20 in sunspot number. The plots show a clear positive correlation between the data, although a close inspection
suggests this correlation to be less clear at the lowest ($\le 30$) and highest ($\ge 150$) sunspot number values. 
As discussed in Section~\ref{intro}, the relation between facular and sunspot area (binned over the sunspot area)  has been described in the literature as either linear or quadratic. In particular, 
\citet{foukal1998} showed that daily data are best represented by a quadratic function, while yearly averaged data are best fitted by a linear relation.
We therefore fitted the 6-month average data and the daily data with both 
analytical functions:

\begin{equation}
A = \alpha + \beta \cdot N_{s}
\label{eq1}
\end{equation}

and
\begin{equation}
A = \alpha + \beta \cdot N_{s} + \gamma \cdot N_{s}^2
\label{eq2}
\end{equation}

where $A$ is the area of the magnetic pixels in $\mu Hem$, $N_{s}$ is the sunspot number and $\alpha$, $\beta$ and $\gamma$ are the free parameters of the fits.
Note that data were fitted between 1996 and 2008 to weigh equally data acquired during different periods of the cycle. The results from the fits are reported in
Table~\ref{tbl-maincoeffs}, together with the values of the Spearman's coefficient $R$ and the result from an F-test. 
Both visual inspection of Figure~\ref{fig2} and the high value found by the F-test clearly indicate that daily data are best fitted by 
a quadratic relation. On the other hand, the fact that the values obtained by the linear and quadratic fits of the 6-months average data for the $\alpha$ and $\beta$ 
parameters agree within the uncertainties of the fits, while the value returned for $\gamma$ is 
smaller than its uncertainty, indicate that the 6-month averaged data are best described by a linear relation. 
This is also confirmed by the small value found for the F-test. \textbf{Note that the value of R, which estimates the monotonicity of the relation between the data points,
is larger than 0.9 for both the daily and the 6-months average data.}

To investigate differences between different phases of the cycle we then analyzed data averaged over running means of 6-months. Results are 
shown in the \textbf{bottom-right} panel of Figure~\ref{fig2}, where the different colors represent data obtained during the ascending (black) and
descending (red) phases of Cycle 23 and during the beginning of Cycle 24 (blue). The plot indicates that for a given sunspot number
the area of magnetic pixels is higher during the descending phase than during the ascending phase of the magnetic cycle. 
Nevertheless, these differences are within the uncertainties of the fit.\newline

\begin{figure}[h]

 \centerline{
               \includegraphics[width=4.9cm,trim=10mm 0 0mm 0, clip=]{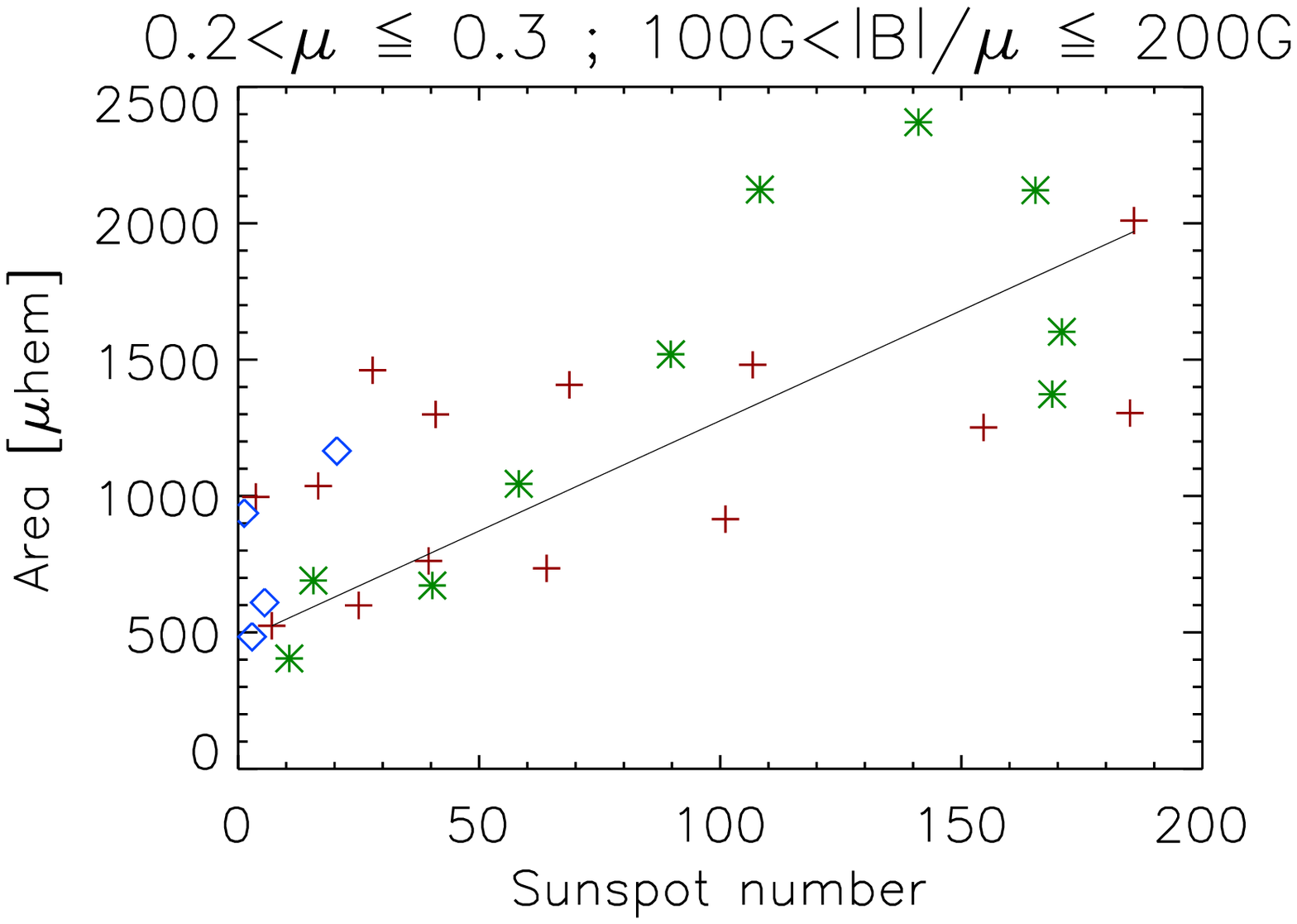}\includegraphics[width=4.9cm,trim=10mm 0 0mm 0,clip=]{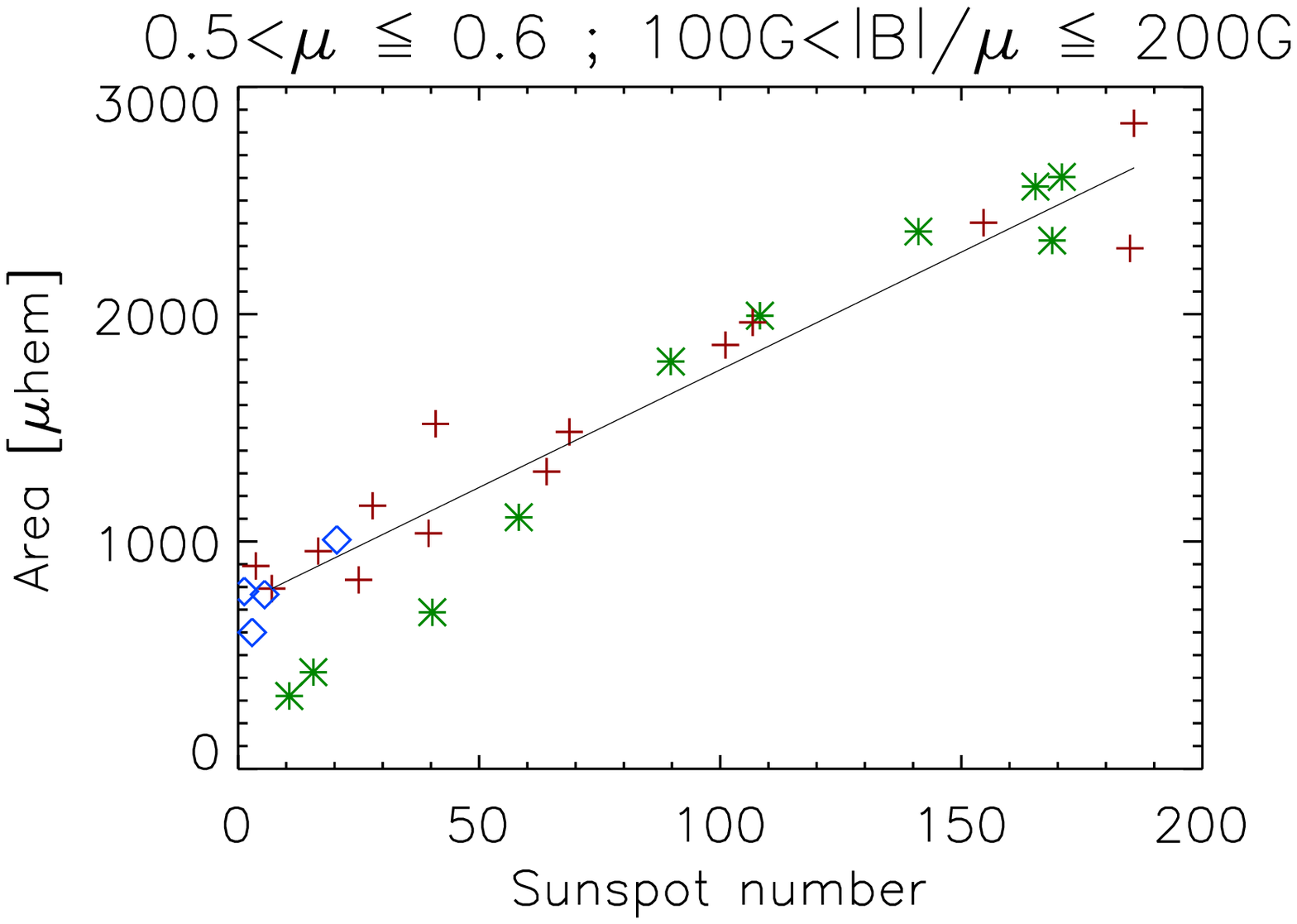}
              \includegraphics[width=4.9cm,trim=10mm 0 0mm 0,clip=]{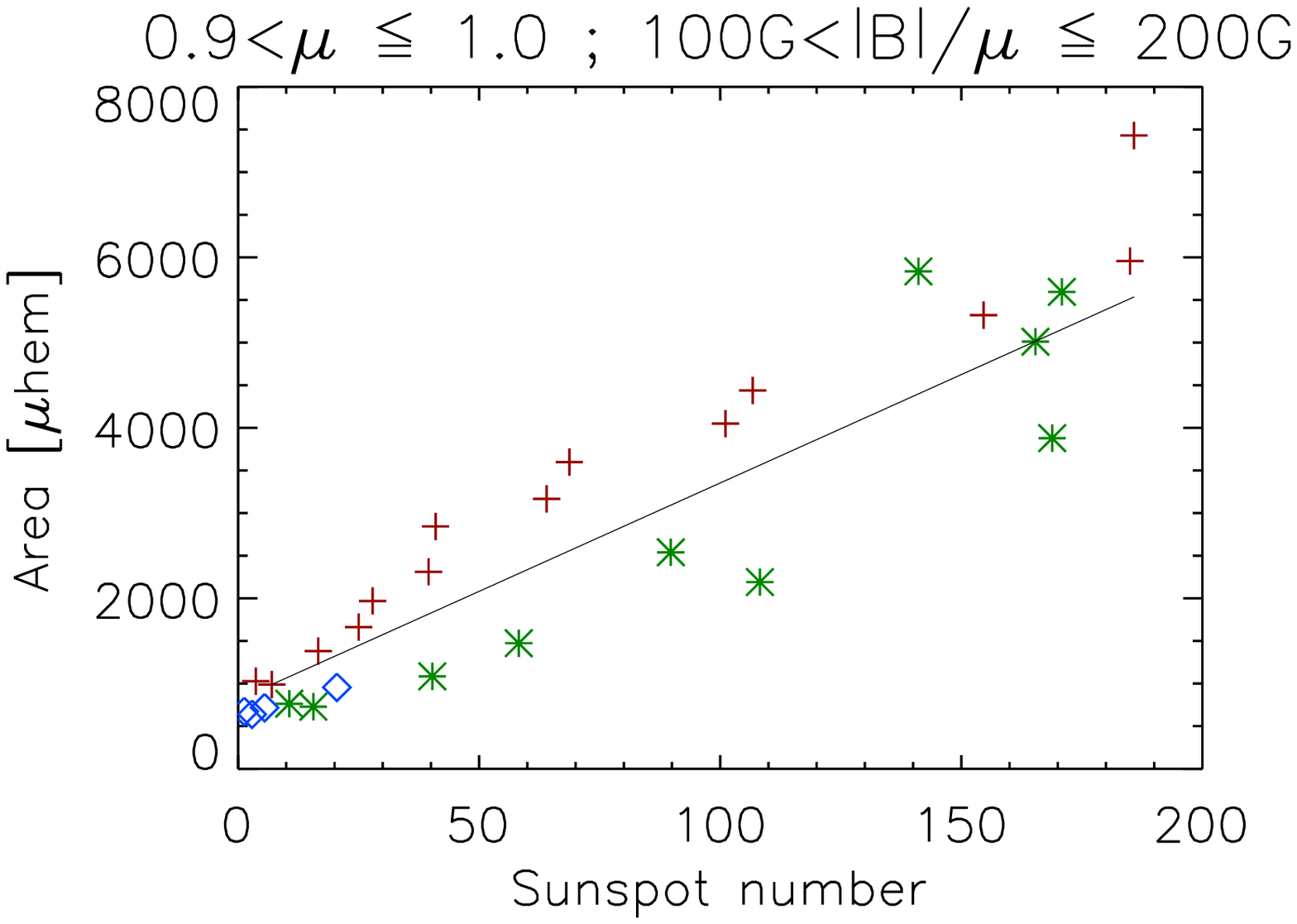}}

 \centerline{
               \includegraphics[width=4.9cm,trim=10mm 0 0mm 0, clip=]{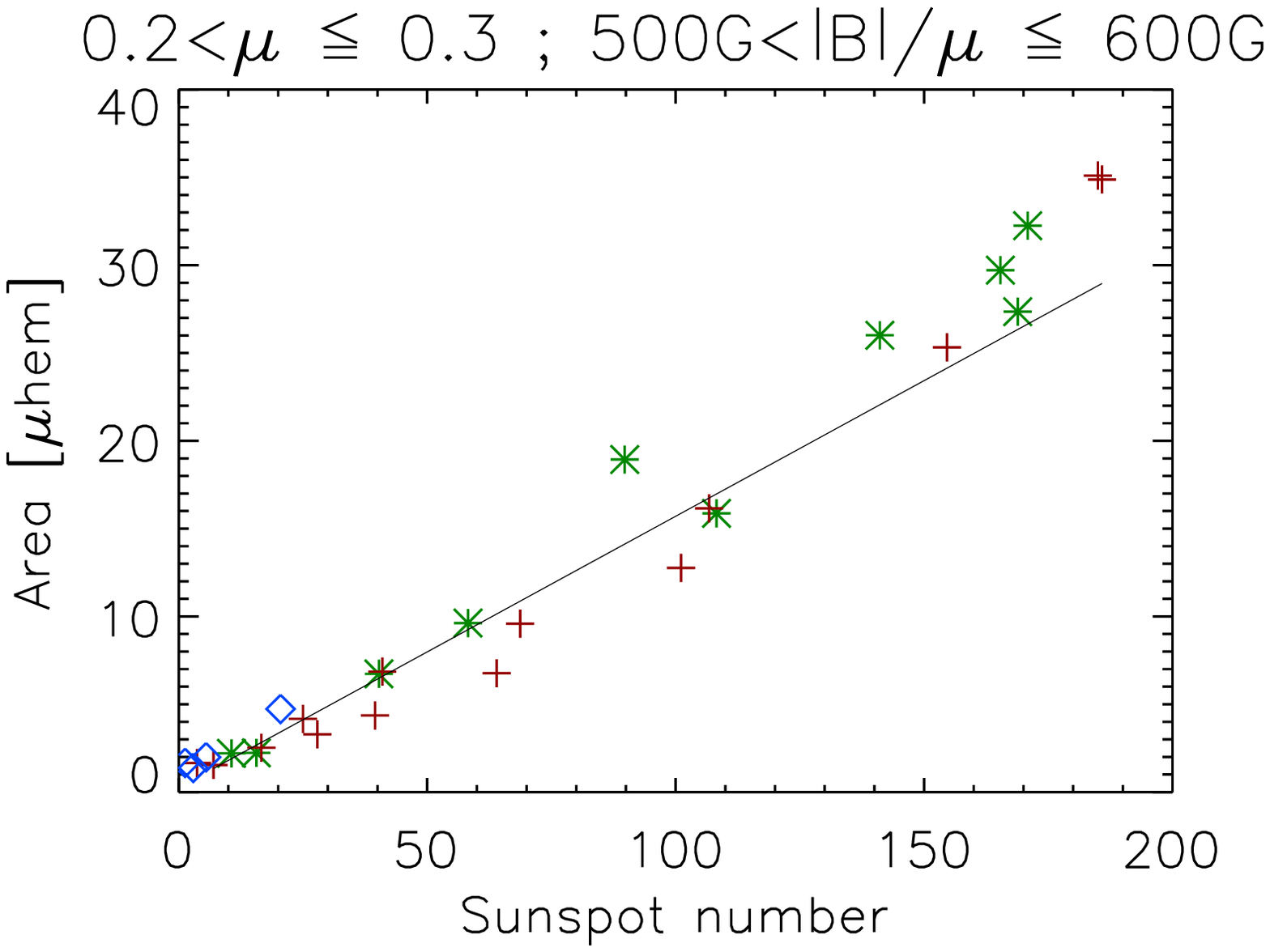}\includegraphics[width=4.9cm,trim=10mm 0 0mm 0,clip=]{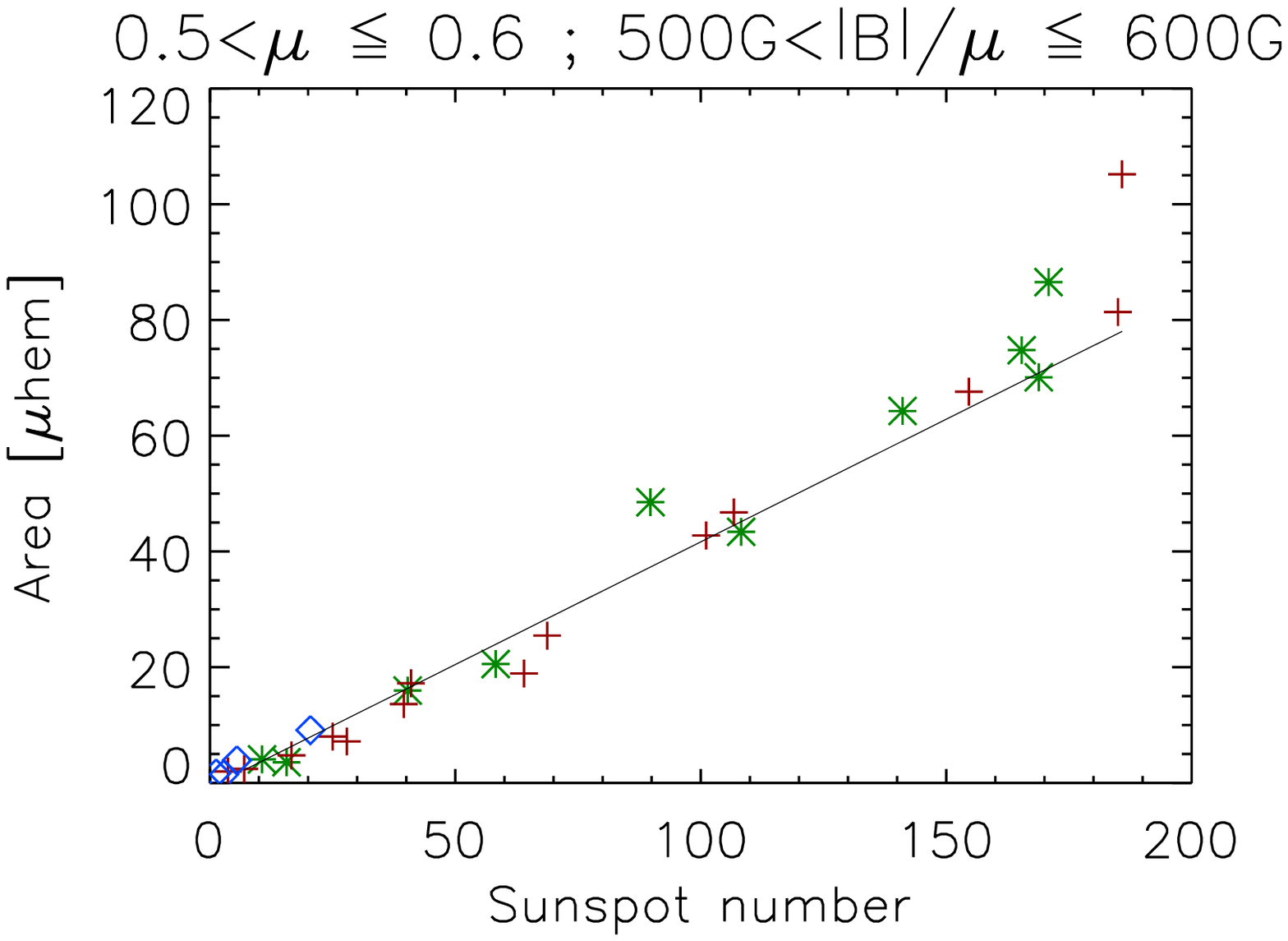}
\includegraphics[width=4.9cm,trim=10mm 0 0mm 0,clip=]{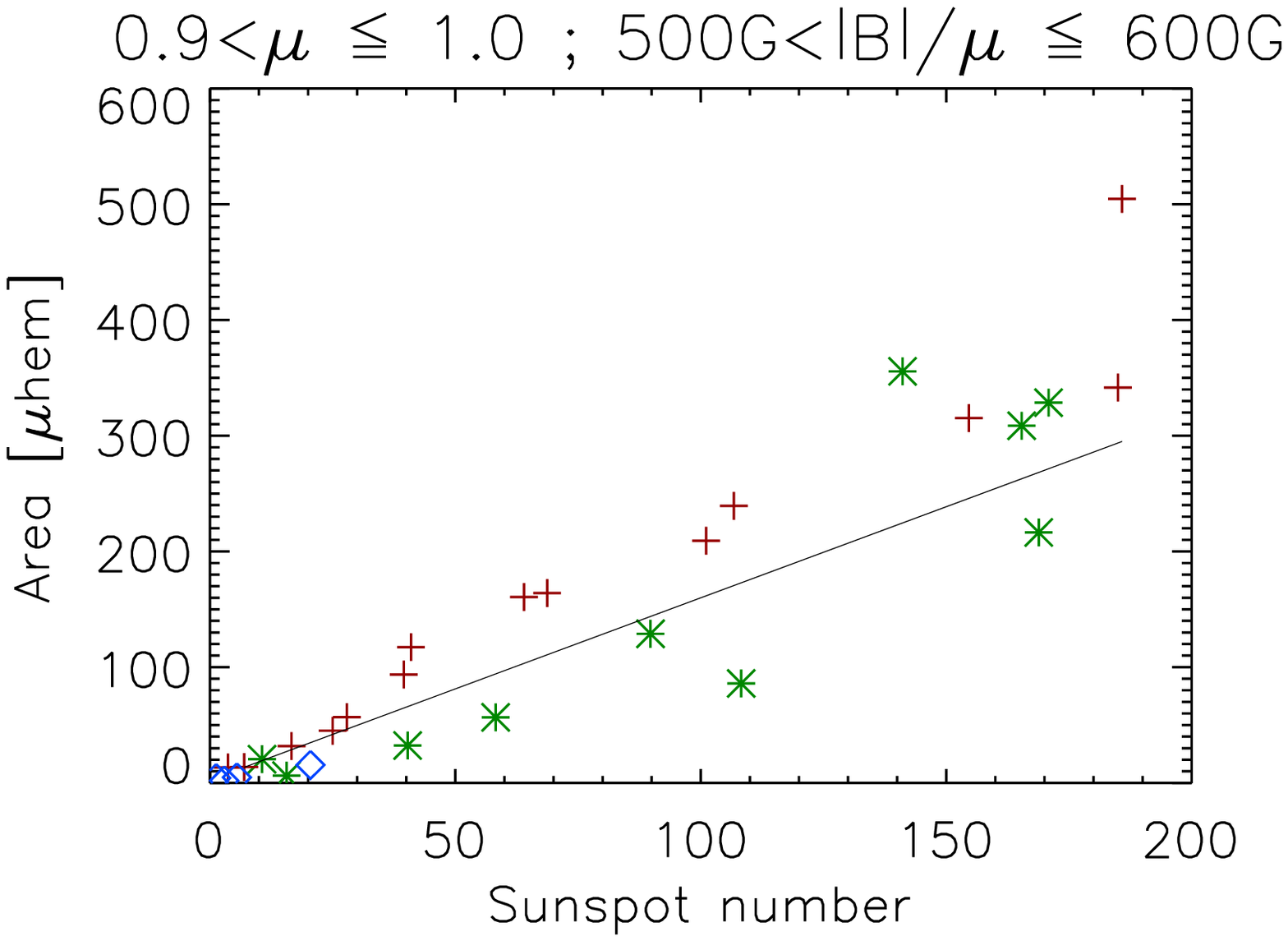}}

 \centerline{
             \includegraphics[width=4.9cm,trim=10mm 0 0mm 0,clip=]{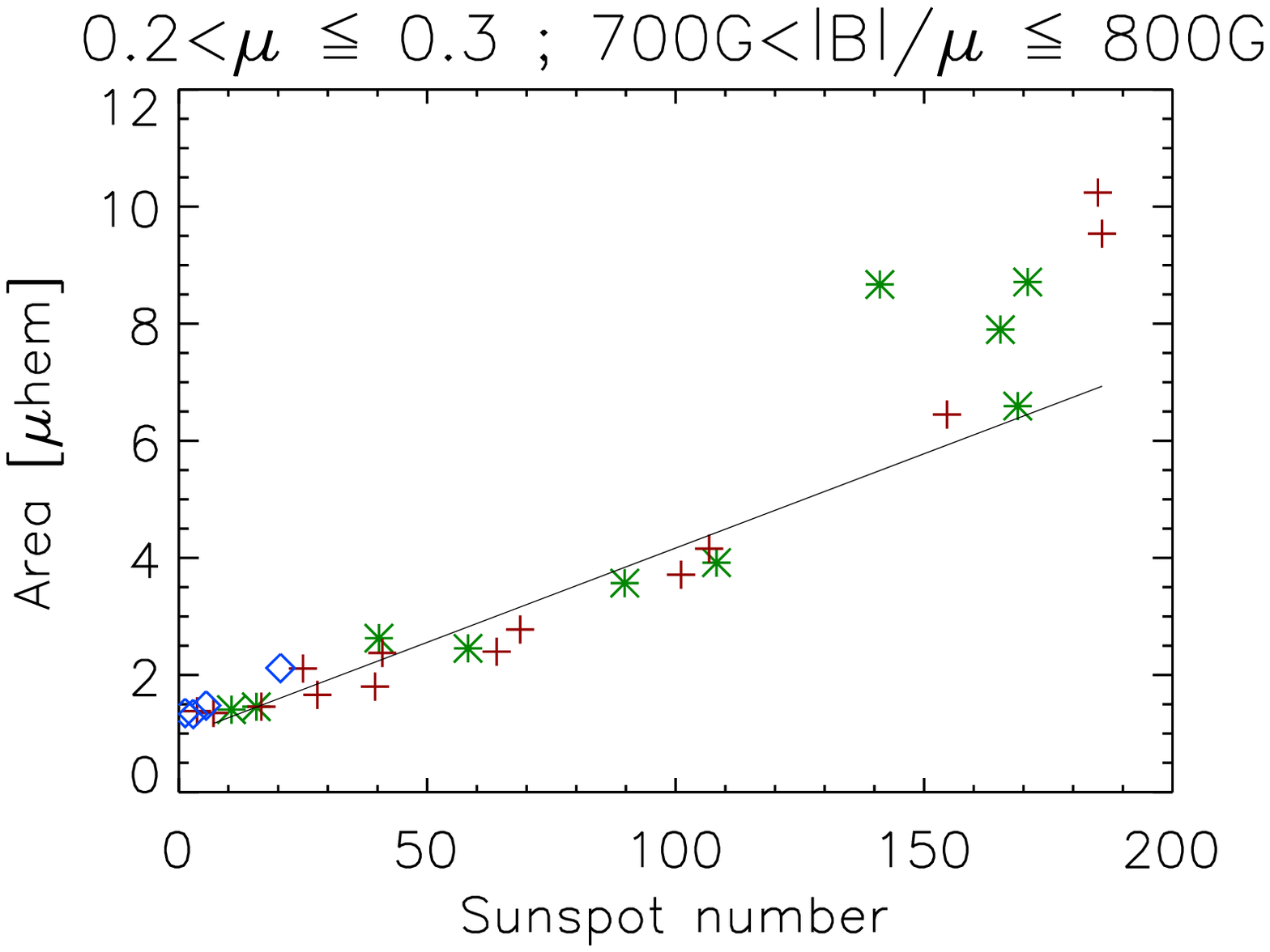}\includegraphics[width=4.9cm,trim=10mm 0 0mm 0,clip=]{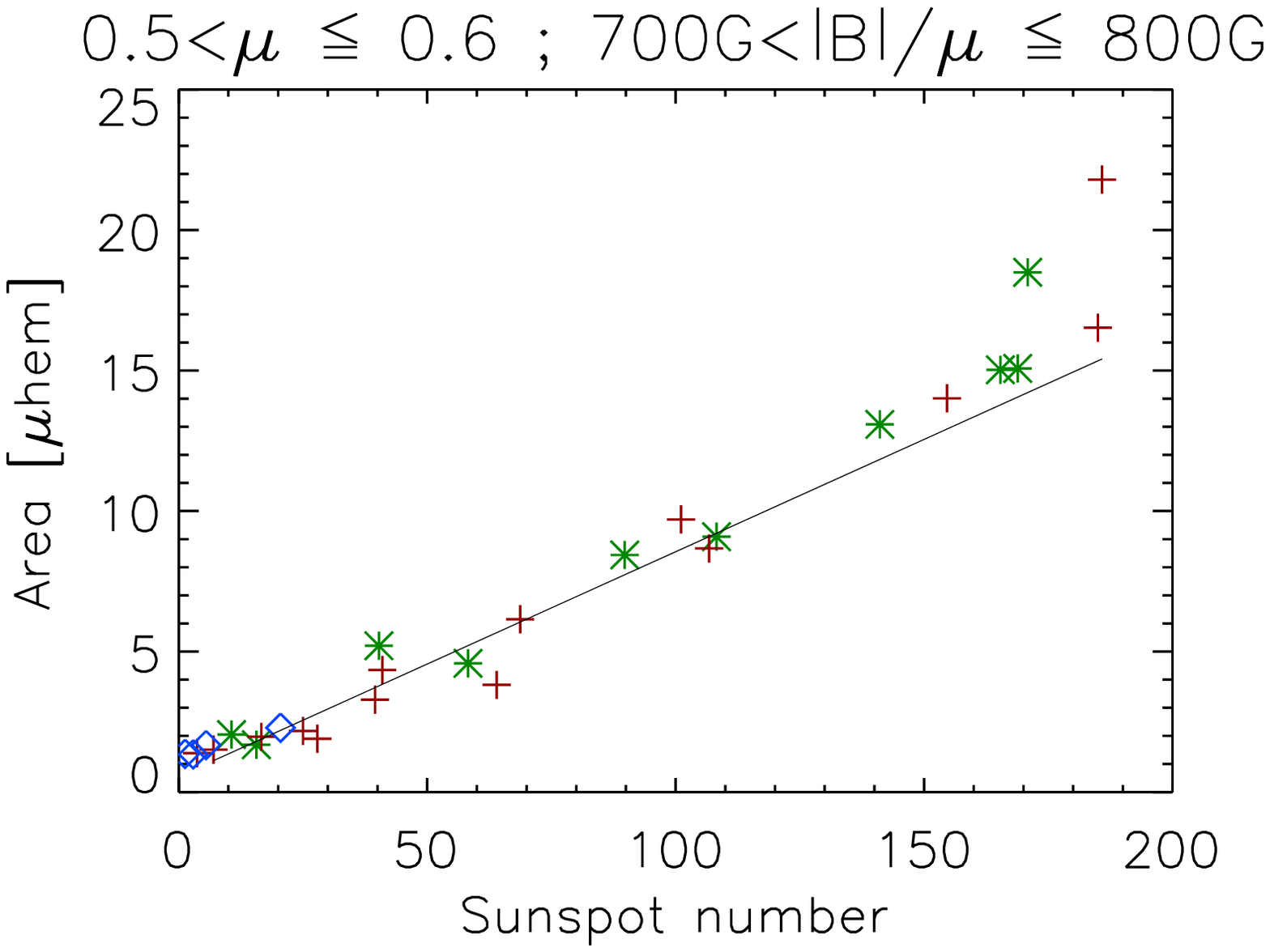}
\includegraphics[width=4.9cm,trim=10mm 0 0mm 0,clip=]{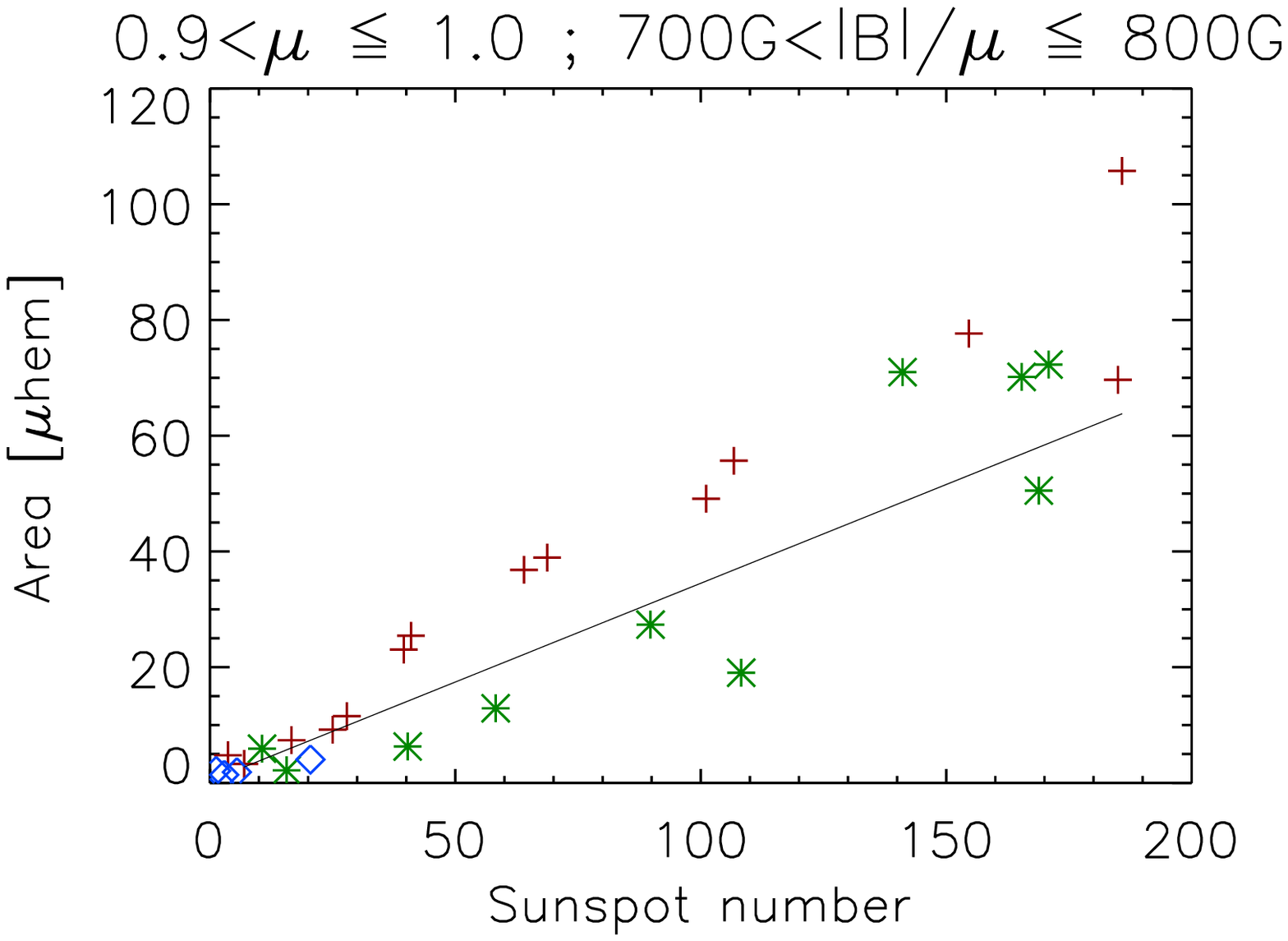}}
\caption{Magnetic pixels area versus sunspot number for pixels located at different positions on the disk and for different magnetic flux values. Green stars: data from
May 1996 to June 2002. Red plus: data from June 2006 to June 2009. Blue diamonds: data from June 2009 to April 2011. Lines represent a linear and a quadratic fit. }
\label{fig3}
\end{figure}

We then repeated the analysis restricting to eight ranges of magnetic flux values and eight positions over the disk. The plots shown in Figure~\ref{fig3} 
report as an example data averaged over 6-months for magnetic pixels located at three different positions over the solar disk,
together with the results
from linear fits. As in Figure~\ref{fig2}, the different colors represent 
data acquired during different phases of the cycle.

We found that in most cases
the data are best represented by a linear relation, the value of the coefficient $\gamma$ being of the order of $10^{-2}$ or smaller, 
and the uncertainty
returned by the fit being typically of the same order or larger than the value of $\gamma$. Only in few cases, typically for the higher 
magnetic flux values and close to disk center, the values of $\gamma$ are larger or of the same order 
as its uncertainty. For these cases the value of $\gamma$ is positive, contrary to what was found when analyzing the daily data, but still of the order of
$10^{-2}$.  We then performed an F-test and 
obtained for all the considered cases a value
larger than the 0.05 confidence level of 4.3. Nevertheless, the small values found for $\gamma$ and inspection of the plots reported in Figure~\ref{fig3}
suggest that the quadratic function best fits only the results obtained during the ascending phase. Such results are therefore not reported.
Results from the linear fits are reported in Table~\ref{tbl-1}. \textbf{Note that in all cases the value of the Spearman's correlation coefficient is larger than 0.9 
(not reported).}

\begin{table}
 \caption{Results from linear fits for different ranges of magnetic flux and positions over the disk of the 6-months averaged data.} 
  \label{tbl-1}
 \resizebox{0.5\textwidth}{!}{  
 \begin{tabular}{c|c|cc}

\hline
$|B|/\mu$ & $\mu$ & $\alpha$ & $\beta$  \\
\hline
          
100-200G  &0.2-0.3&      468$\pm$9&      8$\pm$1 \\
	  &0.3-0.4&      502$\pm$5&      7.9$\pm$0.75\\
	  &0.4-0.5&      724$\pm$6&      8.7$\pm$0.9 \\
	  &0.5-0.6&      720$\pm$7.5&     10$\pm$1  \\
	  &0.6-0.7&      562$\pm$9&      14$\pm$1 \\
	  &0.7-0.8&      520$\pm$12 &      20$\pm$2 \\
	  &0.8-0.9&      610$\pm$15&      26.5$\pm$2\\
	  &0.9-1.&      810$\pm$20&      26$\pm$3 \\
      
      \hline  
200-300G &0.2-0.3&      72$\pm$2&      2.3$\pm$0.3   \\
	&0.3-0.4&      12$\pm$2.5&      2.3$\pm$0.3 \\
	&0.4-0.5&      8$\pm$3&       3.1$\pm$4 \\
	&0.5-0.6&      22$\pm$3.5&      4.1$\pm$0.5 \\
	&0.6-0.7&      30$\pm$4.5&      5.6$\pm$0.6 \\
	&0.7-0.8&      45$\pm$6&      8.1$\pm$0.8 \\
	&0.8-0.9&      70$\pm$8.5&      11$\pm$1 \\
	&0.9-1.&      142$\pm$11&      11$\pm$1.5 \\
   \hline   
300-400G &0.2-0.3&    -0.7$\pm$1.&     0.7$\pm$0.1\\
	&0.3-0.4&    -0.8$\pm$ 1&     0.9$\pm$0.2 \\
	&0.4-0.5&     -2$\pm$2&      1.3$\pm$0.2 \\
	&0.5-0.6&     1$\pm$2&      1.8$\pm$0.3 \\
	&0.6-0.7&    -0$\pm$3&      2.7$\pm$0.4 \\
	&0.7-0.8&     0$\pm$4&      3.9$\pm$0.5 \\
	&0.8-0.9&     1$\pm$5&      5.9$\pm$0.7 \\
	&0.9-1.&      33$\pm$7&      5$\pm$0.9 \\
      \hline  
      
400-500G &0.2-0.3&    -0.$\pm$1&     0.33$\pm$0.07 \\
&0.3-0.4&    0$\pm$1&     0.44$\pm$0.09 \\
&0.4-0.5&     0$\pm$1&     0.60$\pm$0.1 \\
&0.5-0.6&    -0.5$\pm$1&     0.9$\pm$0.1 \\
&0.6-0.7&     -2$\pm$2&      1.3$\pm$0.2 \\
&0.7-0.8&     -2$\pm$2.2&      2.0$\pm$0.3\\
&0.8-0.9&     -6$\pm$7&      3.1$\pm$0.4 \\
&0.9-1.&      9$\pm$4&      3.1$\pm$0.5 \\
      \hline 
      
500-600G &0.2-0.3&     0$\pm$1&     0.15$\pm$0.04 \\
&0.3-0.4&     0$\pm$1&     0.21$\pm$0.05 \\
&0.4-0.5&    0$\pm$1&     0.29$\pm$0.06 \\
&0.5-0.6&    0$\pm$1&     0.42$\pm$0.09 \\
&0.6-0.7&     -1$\pm$1&     0.6$\pm$0.1 \\
&0.7-0.8&     -2$\pm$2&     0.95$\pm$0.1\\
&0.8-0.9&     -3$\pm$3&      1.5$\pm$0.25 \\
&0.9-1.&      2.5$\pm$2.5&      1.6$\pm$0.3 \\
        \hline
      
600-700G &0.2-0.3&  0.7$\pm$0.7&    0.07$\pm$0.02 \\
&0.3-0.4&     0.65$\pm$0.9& 0.1$\pm$0.03 \\
&0.4-0.5&     0.1$\pm$0.9&     0.13$\pm$0.035 \\
&0.5-0.6&     0$\pm$1&     0.19$\pm$0.05 \\
&0.6-0.7&    0$\pm$1&     0.27$\pm$0.06 \\
&0.7-0.8&     -1$\pm$1&     0.42$\pm$0.08 \\
&0.8-0.9&     -2$\pm$1&     0.7$\pm$0.1 \\
&0.9-1.&      1$\pm$1&     0.7$\pm$0.1 \\
      
        \hline
700-800G&0.2-0.3&     0.9$\pm$0.4&    0.03$\pm$0.01 \\
&0.3-0.4&     0.8$\pm$0.6&    0.04$\pm$0.02 \\
&0.4-0.5&     0.50$\pm$0.5&    0.05$\pm$0.02 \\
&0.5-0.6&     0.5$\pm$0.8&    0.08$\pm$0.02 \\
&0.6-0.7&     0.2$\pm$0.8&     0.12$\pm$0.03 \\
&0.7-0.8&    08$\pm$1&     0.18$\pm$0.04 \\
&0.8-0.9&    0$\pm$1&     0.30$\pm$0.06 \\
&0.9-1.&     0$\pm$1&     0.34$\pm$0.08 \\
\hline

 
 \end{tabular}
 }
 
\end{table}

\begin{table}[h]
\caption{Results from linear fits for features located at $0.9< \mu \le 1$ for the ascending (``A'') and descending phases (``D'').}
 \label{tbl-3}
\begin{tabular}{c|c|cc}
 \hline
$|B|/\mu$ & $phase$ &$\alpha$ & $\beta$ \\
\hline            
          
100-200G  &A&       267$\pm$364&      25$\pm$5 \\
          &D&       753$\pm$33&      33$\pm$4 \\
200-300G  &A&       -36$\pm$120&      10$\pm$2\\
          &D&       115$\pm$17&      15$\pm$ 2\\
300-400G  &A&       -40$\pm$60&      5$\pm$1\\ 
	  &D&       20$\pm$10&      8$\pm$1\\
400-500G  &A&        -30$\pm$34&      2.9$\pm$0.8   \\ 
	  &D&       0$\pm$5&      4.3$\pm$0.8  \\ 
500-600G  &A&       -14$\pm$14&      1.5$\pm$0.4    \\
          &D&       -2$\pm$3&      2.3$\pm$0.5   \\
600-700G  &A&       -5$\pm$5&     0.7$\pm$0.2   \\
          &D&       0$\pm$2&     1.1$\pm$0.2   \\
700-800G  &A&       -3$\pm$3&     0.3$\pm$0.1  \\
          &D&       0$\pm$1&     0.5$\pm$0.1   \\

  \hline
         \end{tabular}

\end{table}

\begin{figure}
 \centerline{
               \includegraphics[width=0.55\textwidth,clip=]{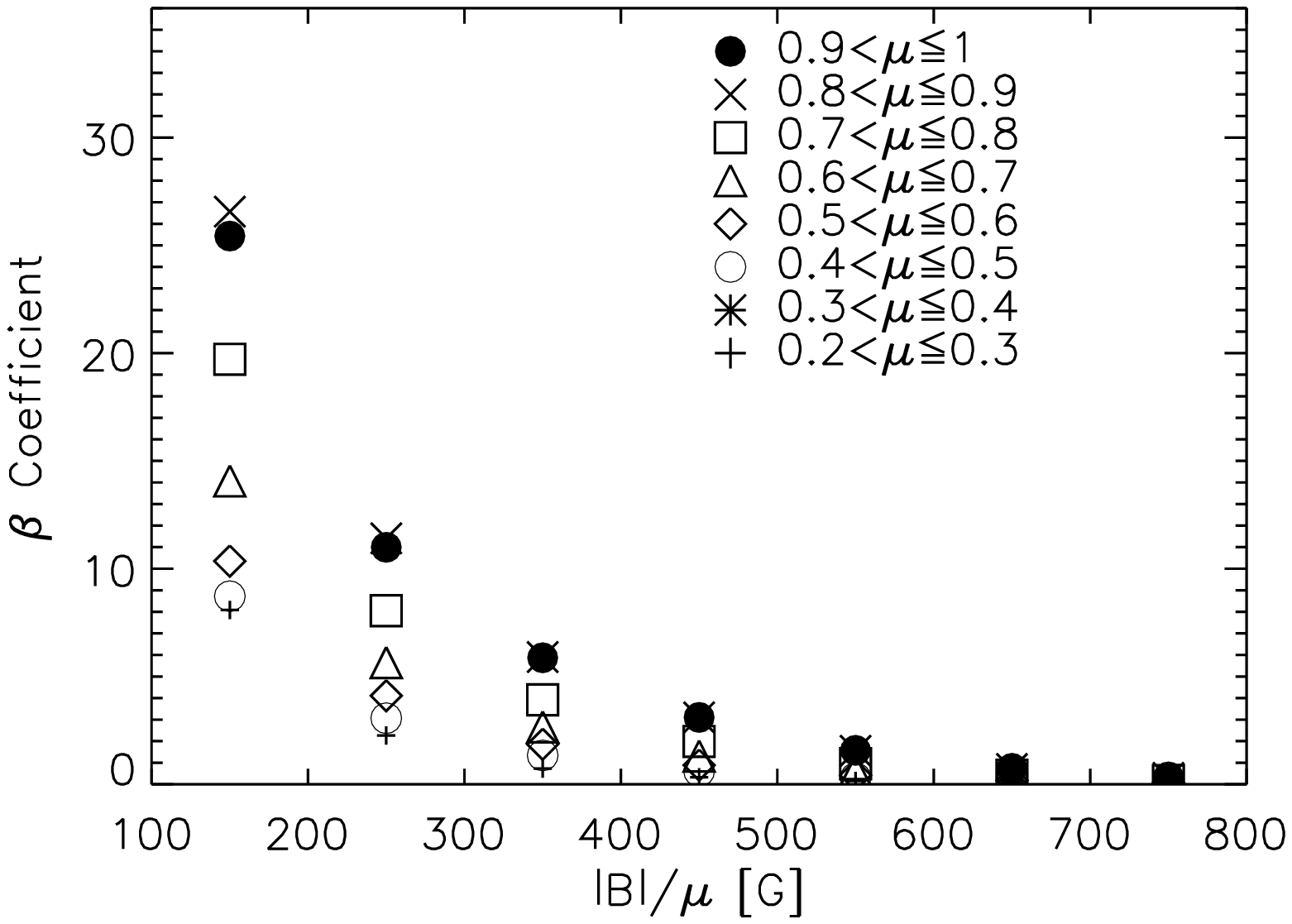}\includegraphics[width=0.55\textwidth,clip=]{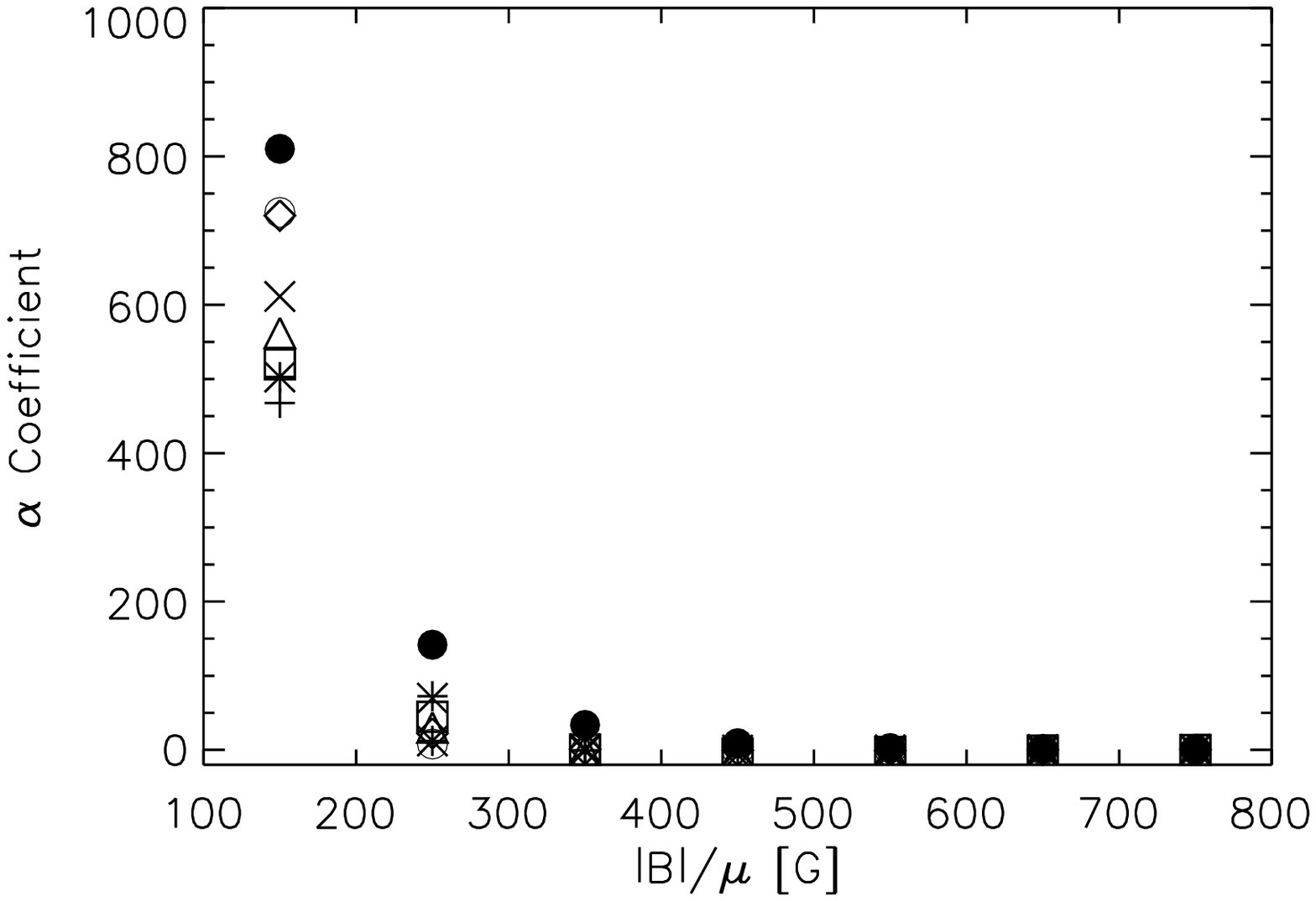}
}

\caption{Variation of the linear fit coefficients of the 6-months averaged data with the magnetic flux derived for different positions over the disk. Data-points values are reported in 
Table \ref{tbl-1}.}
\label{fig4}
\end{figure}

Analysis of the results reported in Table~\ref{tbl-1} reveals that the values of the $\alpha$  and $\beta$ coefficients of the linear relation are a function of both
the position on the disk $\mu$ and
the magnetic flux. This dependence is illustrated in plots in Figure~\ref{fig4}, that show the variation of both $\beta$ and $\alpha$ with the increase 
of the magnetic flux for 
different values of $\mu$. The values of $\beta$ and  $\alpha$ decrease with the increase of the magnetic flux; in particular, 
it is worth to note that $\alpha$ is significantly different from zero only for $|B|/\mu \le 300G$. The value of both coefficients shows instead an overall 
increase with $\mu$, although the dependence of $\alpha$ on the position over the disk is not strictly monotonous, most likely because of noise effects 
(see Sec.~\ref{Sec.Noi}).

The plots in Figure~\ref{fig3} show that, as also found when analyzing the whole dataset, especially close to disk center for a given sunspot number
the area of magnetic pixels is higher during the descending phase then 
during the ascending phase of the magnetic cycle. In this case, the difference is not negligible as the area of magnetic features during the descending phase 
can be twice the one 
measured  during the ascending phase and it is larger than the uncertainties derived by the fits. 
In Table~\ref{tbl-3} we report results from linear fits to data acquired at disk center
($0.9 < \mu \leq 1.$) during 
the ascending (from 1996 to 2002) and descending phase (from 2002 to 2009), separately. Even in this case second order polynomial fits to the data do not return 
statistically significant results, while the slopes computed during the descending phase are systematically higher than the slopes computed during the ascending phase.  
\newline

\begin{figure}

 \centerline{
               \includegraphics[width=4.9cm,trim=10mm 0 0mm 0, clip=]{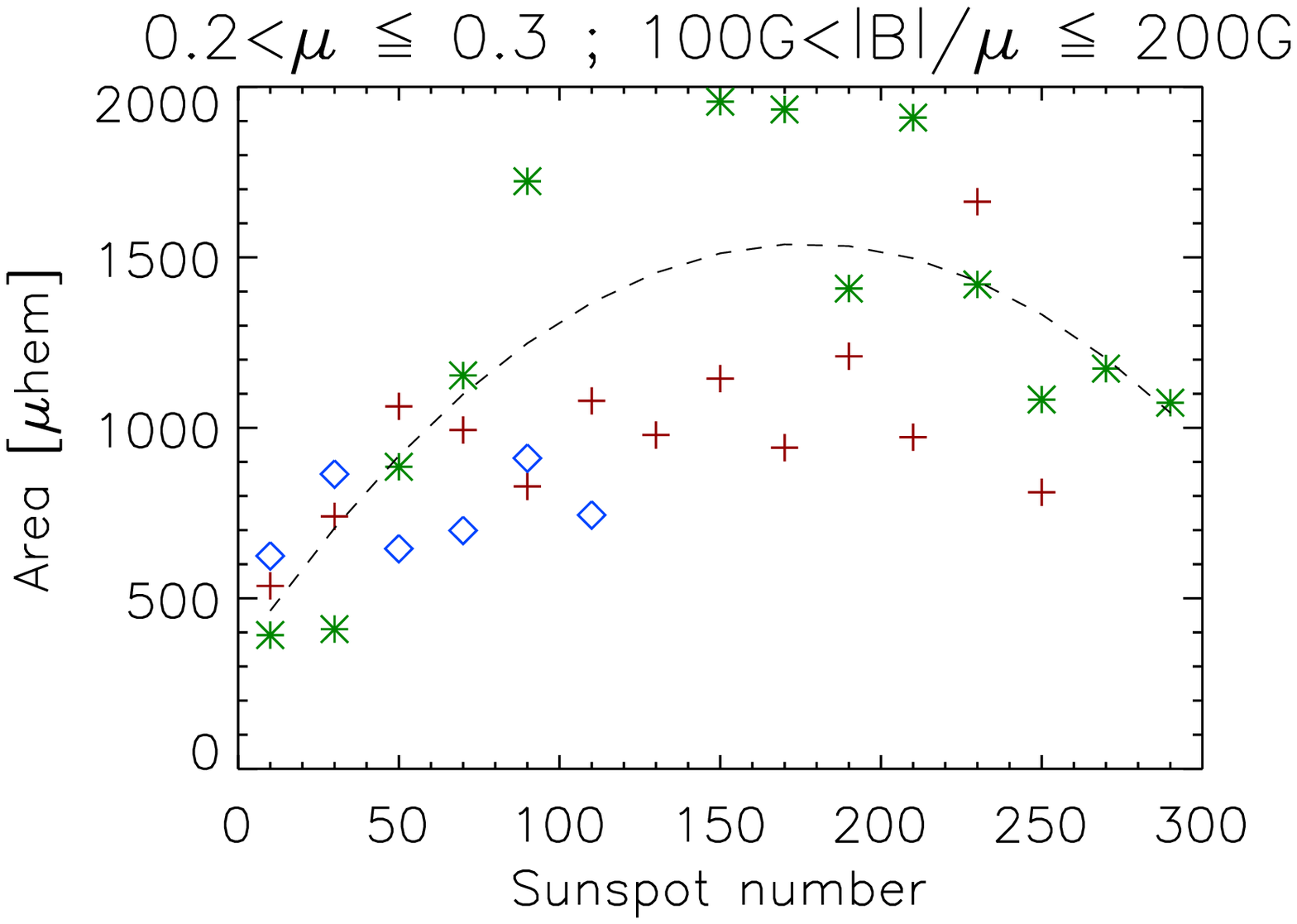}\includegraphics[width=4.9cm,trim=10mm 0 0mm 0,clip=]{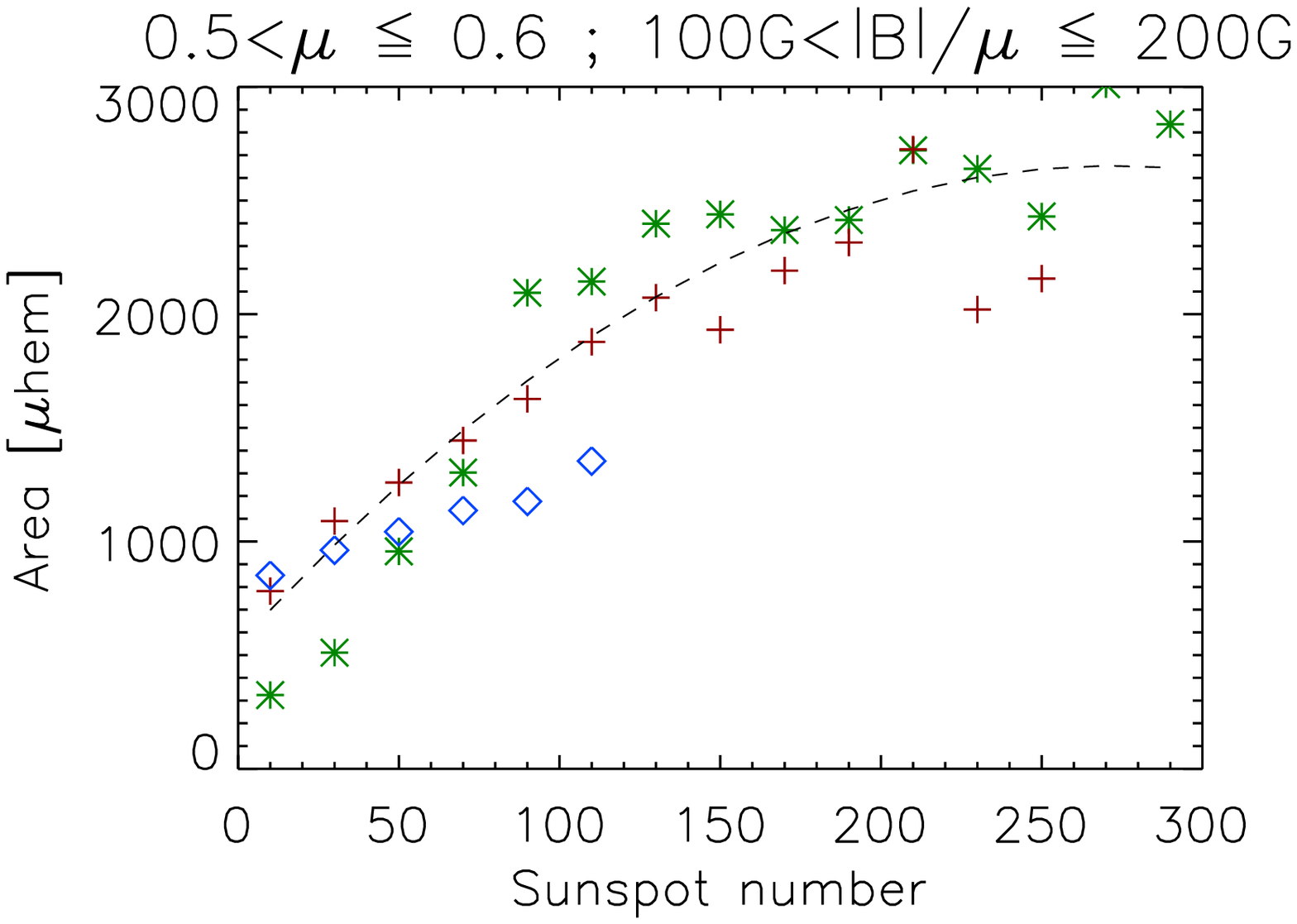}
              \includegraphics[width=4.9cm,trim=10mm 0 0mm 0,clip=]{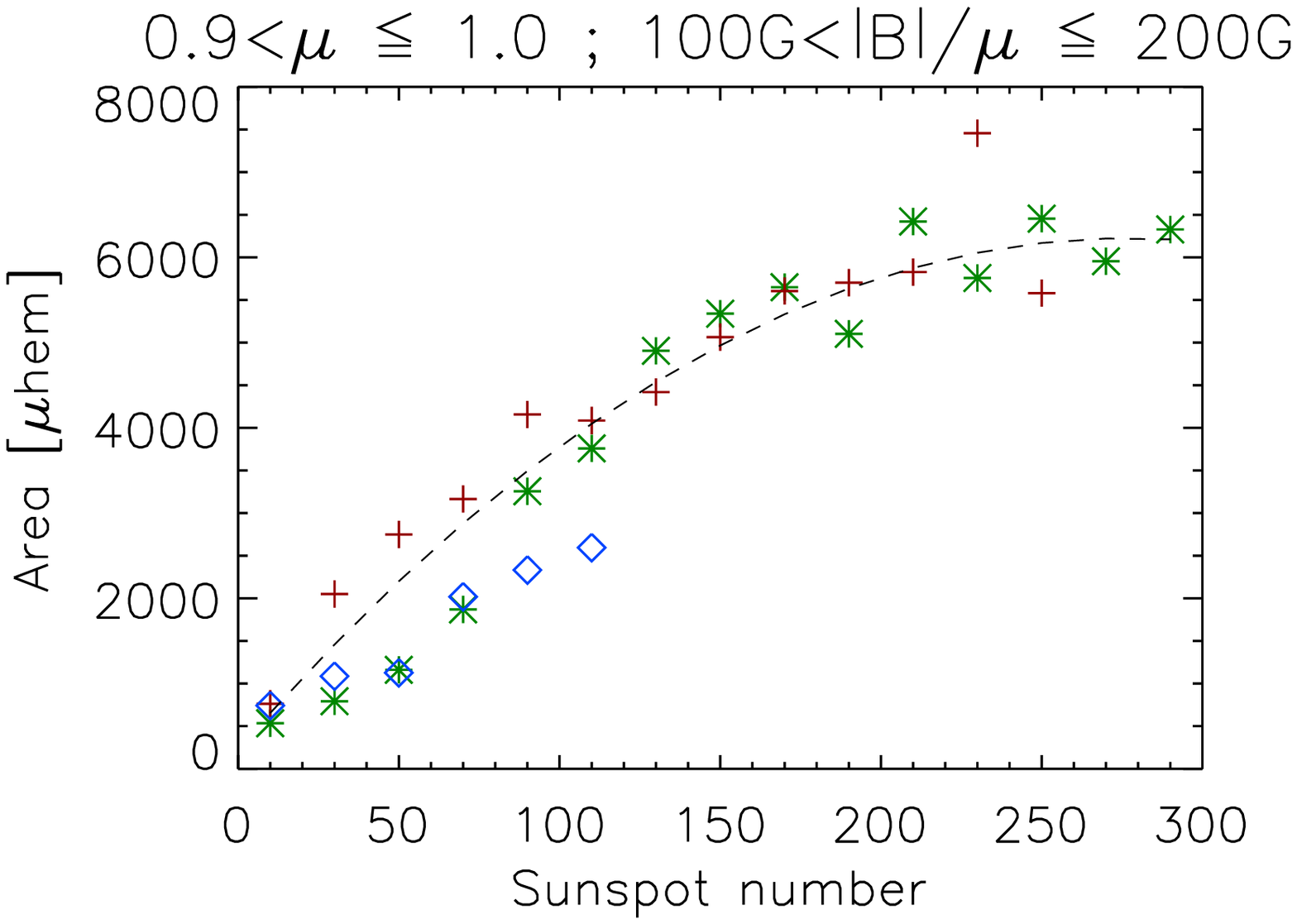}}

 \centerline{
               \includegraphics[width=4.9cm,trim=10mm 0 0mm 0, clip=]{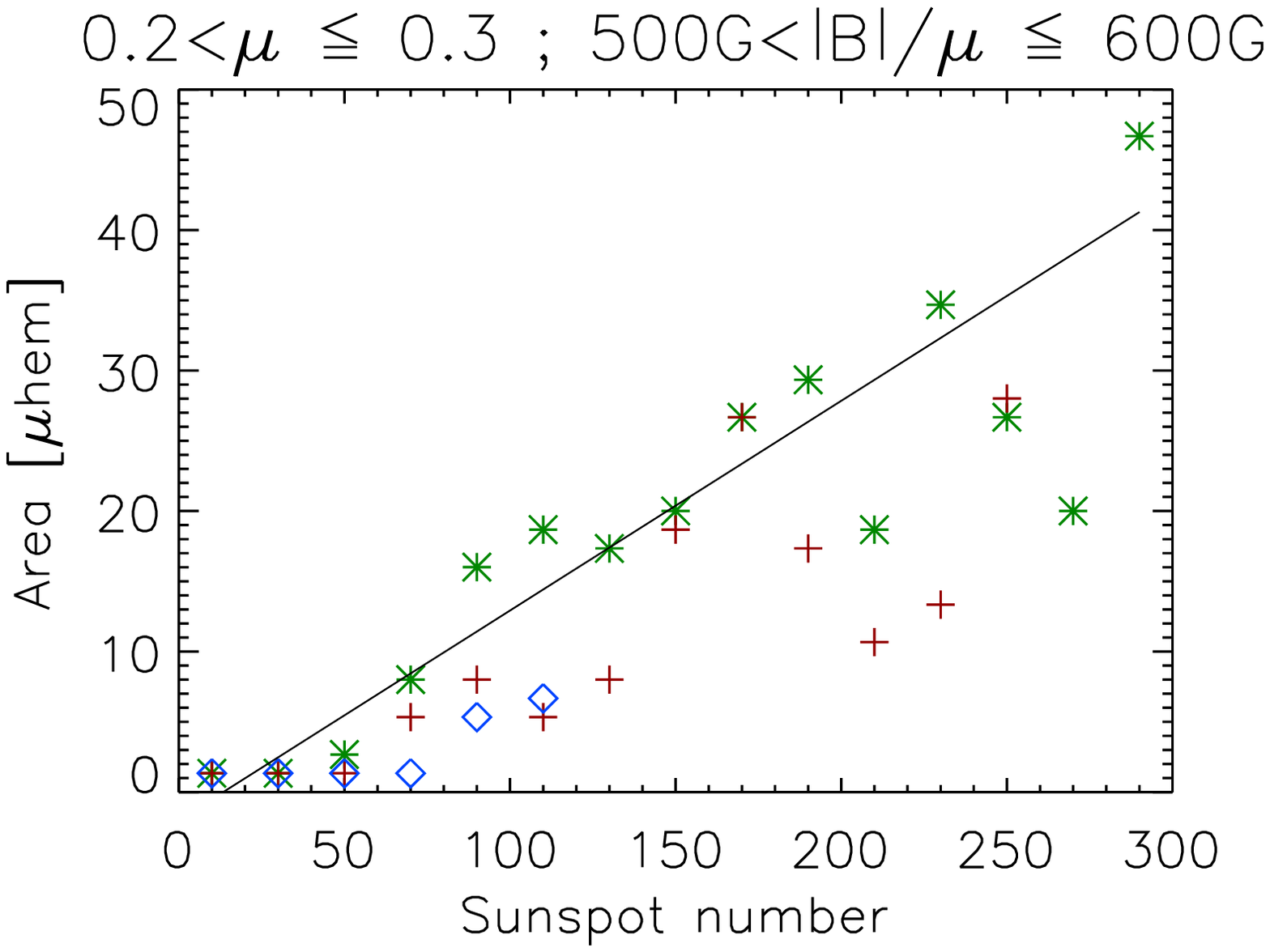}\includegraphics[width=4.9cm,trim=10mm 0 0mm 0,clip=]{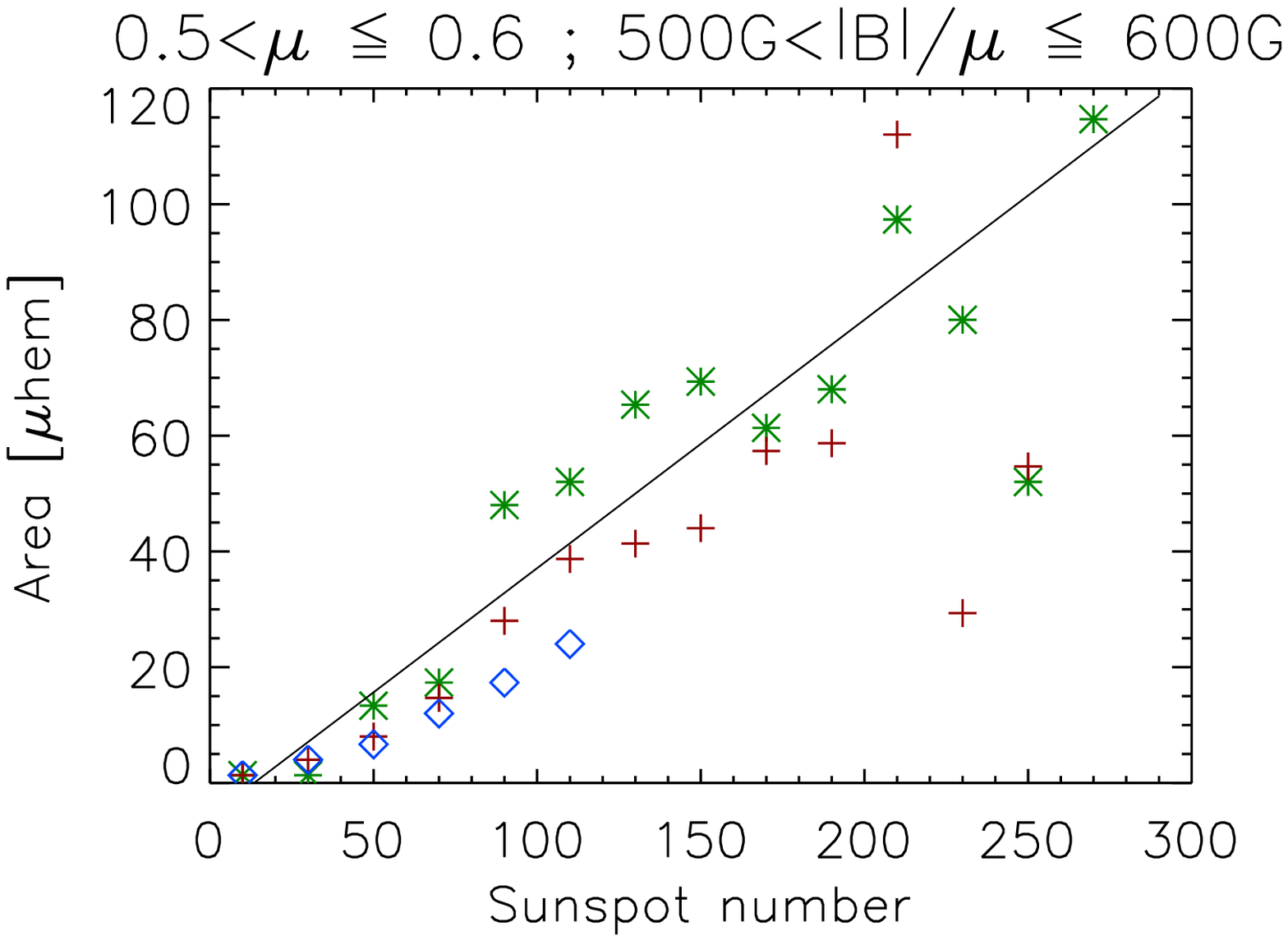}
\includegraphics[width=4.9cm,trim=10mm 0 0mm 0,clip=]{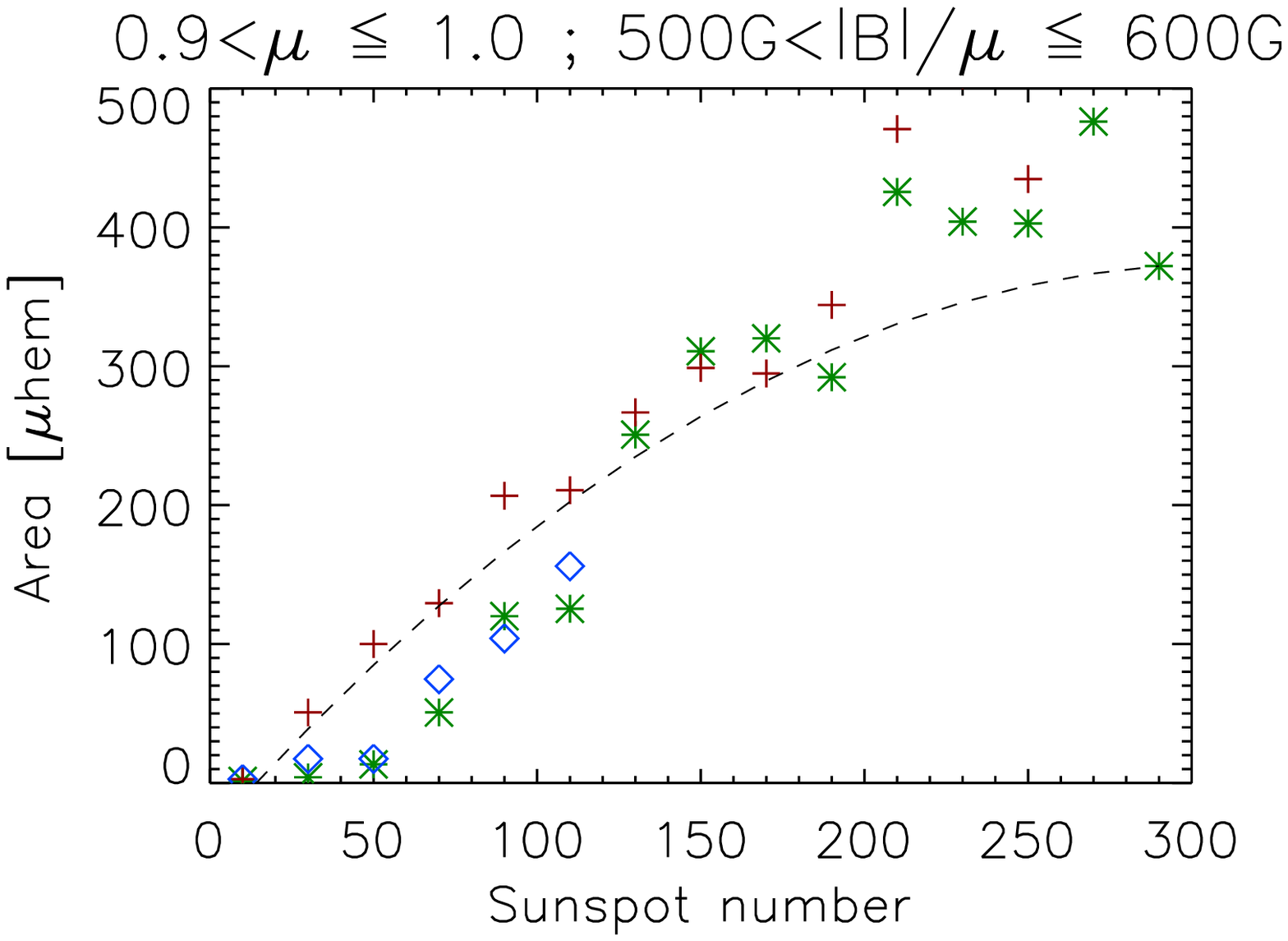}}

 \centerline{
             \includegraphics[width=4.9cm,trim=10mm 0 0mm 0,clip=]{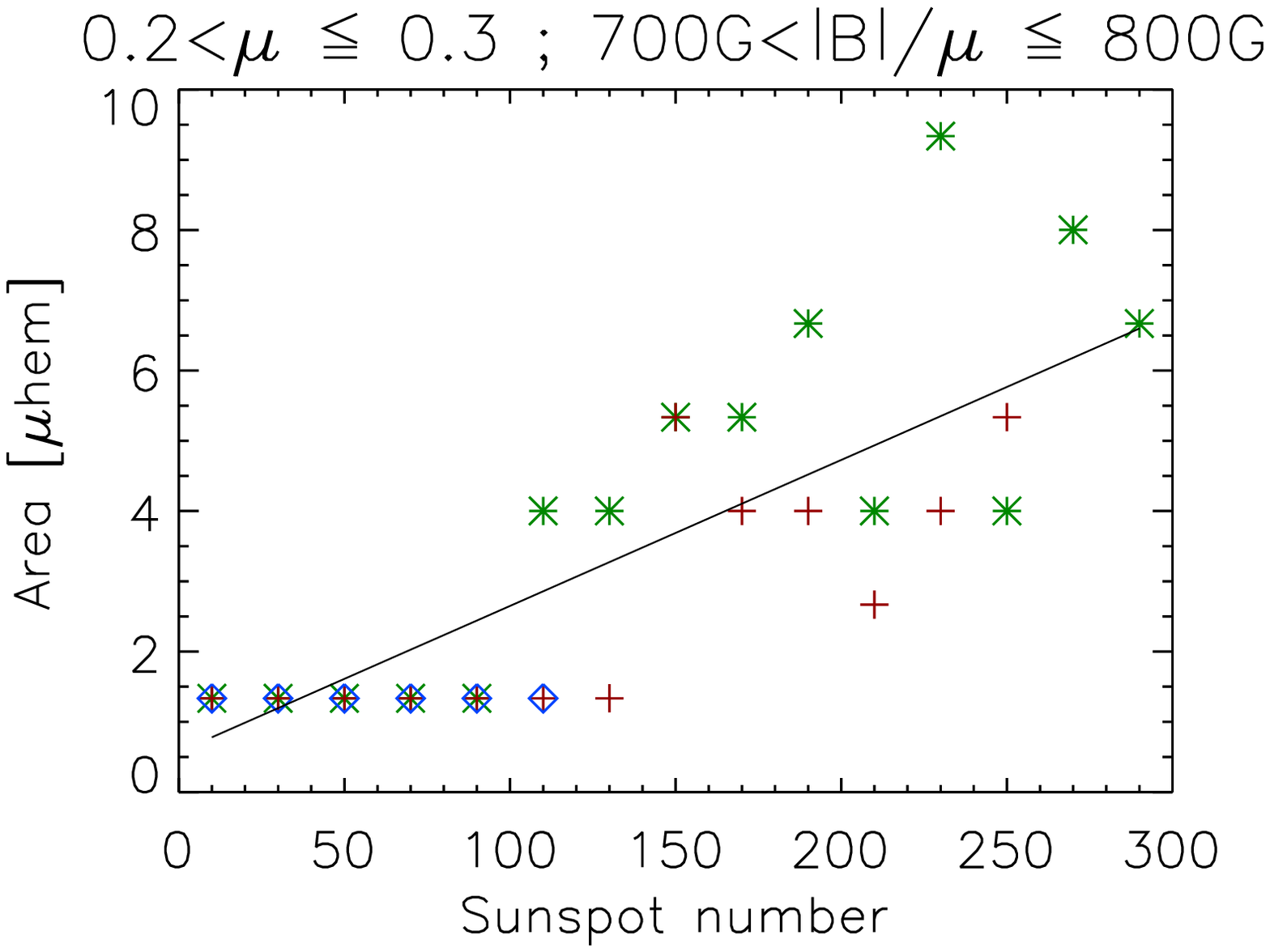}\includegraphics[width=4.9cm,trim=10mm 0 0mm 0,clip=]{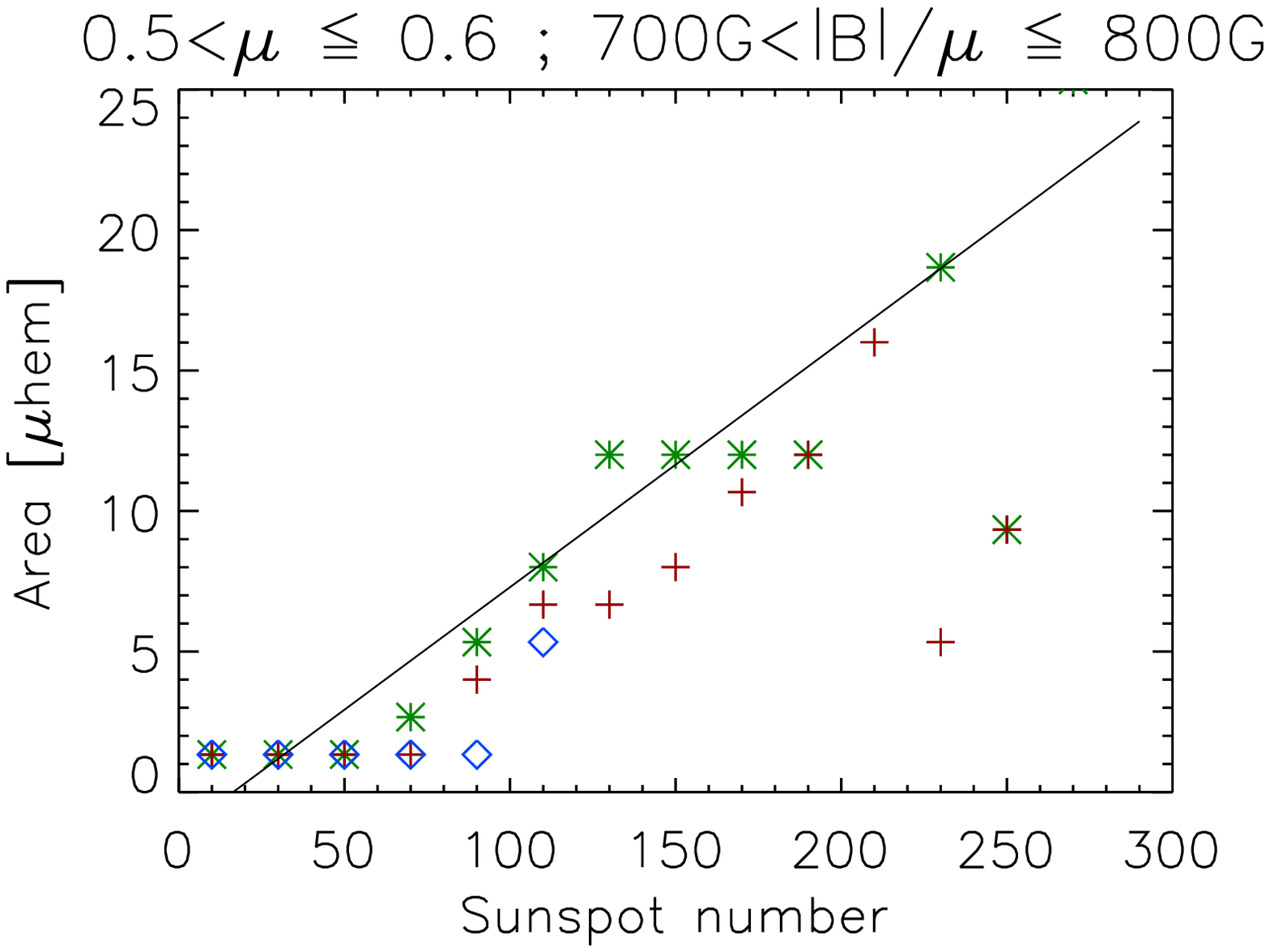}
\includegraphics[width=4.9cm,trim=10mm 0 0mm 0,clip=]{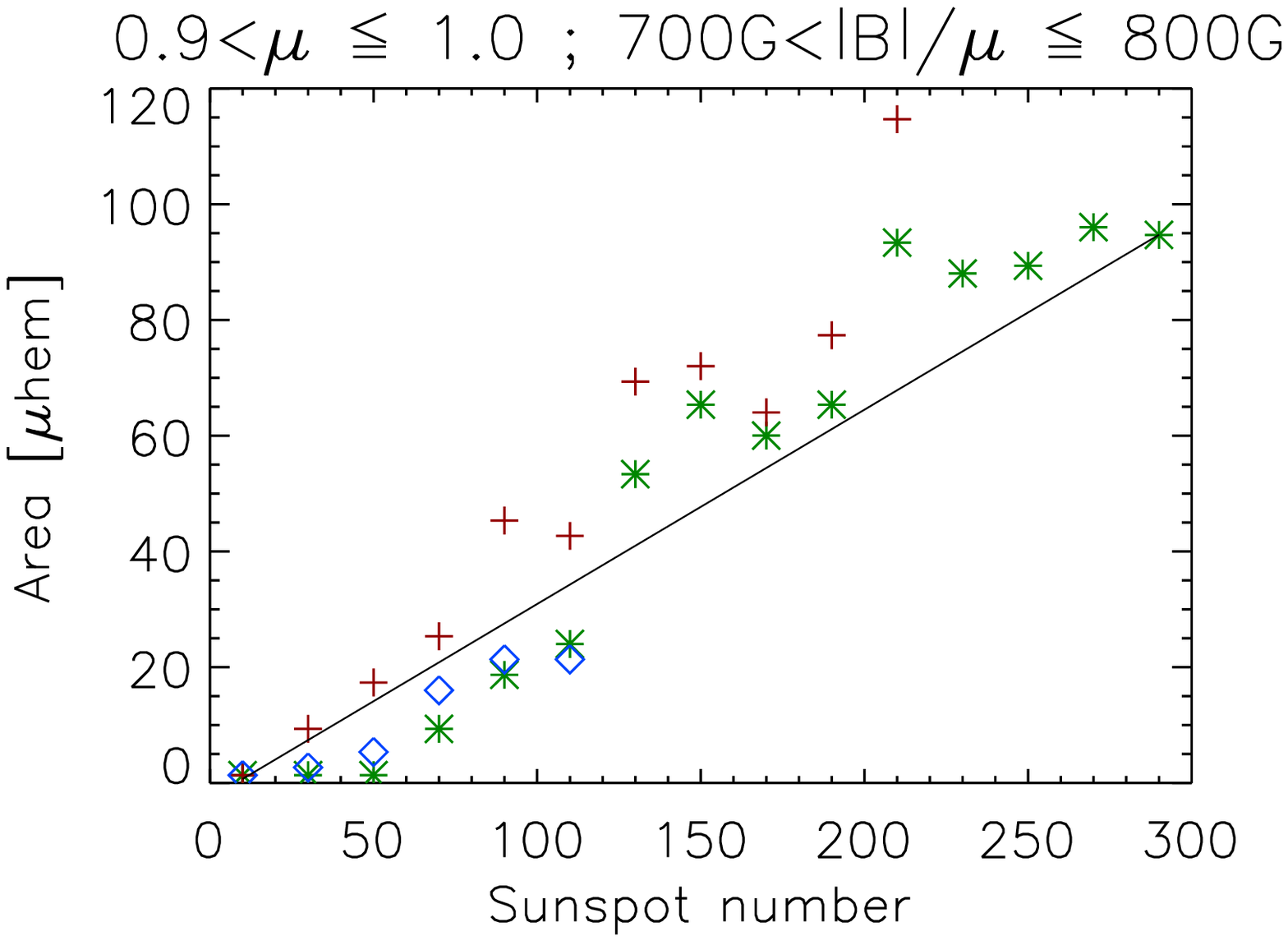}}
\caption{Magnetic pixels area versus sunspot number for pixels located at different positions on the disk and for different magnetic flux values for 
data binned over the sunspot number. Green stars: data from
May 1996 to June 2002. Red plus: data from June 2006 to June 2009. Blue diamonds: data from June 2009 to April 2011. Lines represent a linear and a quadratic fit. }
\label{fig31}
\end{figure}

Linear and quadratic fits were then performed on daily data.
\textbf{For these data the value of the Spearman's coefficient, with the exception of few cases, was $\geq$ 0.8.}
We found that, in general, for magnetic flux values larger than 400G the value returned 
for $\gamma$ by the second order fit is smaller than its uncertainty and the F-test returns a value smaller than the 4.3 threshold confidence level, so that 
 these cases are
best 
represented by a linear function, while for magnetic fluxes smaller than 400G data are best represented by a quadratic relation with $\gamma < 0$.
Figure~\ref{fig31} shows 
for instance the scatter-plots together with the corresponding best-fits for the same cases illustrated in Figure~\ref{fig3}. In Table~\ref{tbl-4} we report for each 
position over the disk and for each magnetic range considered only the results obtained for the analytical function that best fits the data, together with the F-test value.
We note that, in agreement with results obtained for 6-month averaged data, the values of $\alpha$ and $\beta$ decrease with the increase of
the magnetic flux, and that the same trend is found for the absolute value of $\gamma$. An easy analysis of the values reported in the table shows 
that the maxima of the quadratic
fits shift toward higher sunspot number values with the increase of the magnetic flux, thus suggesting that most likely for magnetic flux values larger
than 400G the data are best represented 
by a linear relation because we could not observe the maximum of the relation between
sunspot number and magnetic pixels area. The data reported in Table~\ref{tbl-4} also show that the values of $\beta$ 
and the absolute value of $\gamma$ increase from the limb toward disk center, while no clear trend is found for $\alpha$.

Finally, the plots in Figure~\ref{fig31} show no clear difference between results obtained during different phases of the cycle.

\begin{table}
 \caption{Results from linear and quadratic fits for different ranges of magnetic flux and positions over the disk obtained for the daily data.} 
  \label{tbl-4}
 \resizebox{0.75\textwidth}{!}{  
 \begin{tabular}{c|c|ccc|c}

\hline
$|B|/\mu$ & $\mu$ & $\alpha$ & $\beta$ & $\gamma$ & F-test \\
\hline
          
100-200G &0.2-0.3& 330$\pm$400&      14$\pm$7     &  -0.04$\pm$0.02&      67\\
	 &0.3-0.4& 470$\pm$190&      9$\pm$3  & -0.015$\pm$0.01&      14\\
	&0.4-0.5&  584$\pm$251&      13$\pm$4 &   -0.025$\pm$0.01&      60\\
	&0.5-0.6&  546$\pm$294&      15$\pm$5 &  -0.03$\pm$0.01&      65\\
	&0.6-0.7&  357$\pm$283&      21$\pm$5 &   -0.04$\pm$0.02&      169\\
	&0.7-0.8&  254$\pm$340&      29$\pm$7 &   -0.06$\pm$0.02&      466\\
	&0.8-0.9&  257$\pm$436&      40$\pm$9 &   -0.07$\pm$0.03&      477\\
	&0.9-1.&   229$\pm$625&      43$\pm$13 &   -0.08$\pm$0.04&      209\\
      
      \hline  
200-300G &0.2-0.3&      51$\pm$52&      2.6$\pm$1.&  -0.004$\pm$   0.003&      43\\
	&0.3-0.4&     -14$\pm$45&      3$\pm$1&  -0.004$\pm$0.0035&      46\\
	&0.4-0.5&     -21$\pm$60&      3.6$\pm$1.4 &  -0.004$\pm$0.004&      10\\
	&0.5-0.6&     -30$\pm$80&      5.4$\pm$1.8 &  -0.008$\pm$0.005&      28\\
	&0.6-0.7&     -45$\pm$93&      8$\pm$2&  -0.014$\pm$0.007&      135\\
	&0.7-0.8&     -63$\pm$136&      11$\pm$3&   -0.021$\pm$0.009&      234\\
	&0.8-0.9&     -61$\pm$180&      15$\pm$4&   -0.02$\pm$0.01&      92\\
	&0.9-1.&     -117$\pm$260&      19$\pm$6 & -0.03$\pm$0.02&      141\\
      
   \hline   
300-400G &0.2-0.3&      -3$\pm$16&     0.54$\pm$0.05& - &     2 \\
      &0.3-0.4&     -12$\pm$25&      1.0$\pm$0.6 &  -0.001$\pm$0.001&      14\\
      &0.4-0.5&     -19$\pm$27&     1.7$\pm$0.7 &  -0.0025$\pm$0.002&      22\\
      &0.5-0.6&     0$\pm$33&      1.4$\pm$0.1& - &     4\\
      &0.6-0.7&     -38$\pm$46&      3.7$\pm$1.2 & -0.0065$\pm$0.004&      58\\
      &0.7-0.8&     -51$\pm$73&    5.3$\pm$1.7 &  -0.0085$\pm$0.005&      115\\
      &0.8-0.9&     -62$\pm$104&     7.35$\pm$2. &  -0.009$\pm$0.007&      33\\
      &0.9-1.&     -120$\pm$162 &   10.$\pm$3.5 &  -0.02$\pm$0.01&      75\\
      \hline  
      
400-500G& 0.2-0.3&   -2$\pm$4&     0.15$\pm$0.01& - &     0.6\\
	  &0.3-0.4&   -3$\pm$5&     0.19$\pm$0.02& - &    2\\
	  &0.4-0.5&  -8$\pm$9&     0.31$\pm$0.03& - &   3\\
	  &0.5-0.6&   -6$\pm$7&     0.43$\pm$0.02& - & 1.5\\
	&0.6-0.7&     -22$\pm$24&      1.8$\pm$0.7 & -0.003$\pm$0.002&      43\\
	&0.7-0.8&     -42$\pm$47&      2.9$\pm$1.0 & -0.005$\pm$0.003&      65\\
	&0.8-0.9&     -48$\pm$69&      4$\pm$1.5 &  -0.004$\pm$0.004&      38\\
	&0.9-1.&     -56$\pm$68&      5$\pm$2 & -0.008$\pm$0.006&      49\\
      
      \hline 
      
500-600G &0.2-0.3& -2.$\pm$4.&     0.15$\pm$0.01& - &   3.9 \\
&0.3-0.4&     -3$\pm$5&     0.19$\pm$0.02& - &    1.\\
&0.4-0.5&     -8$\pm$9&     0.31$\pm$0.0& - &    0.6\\
&0.5-0.6&     -6$\pm$7&     0.43$\pm$0.02& -&   0.6 \\
&0.6-0.7&     -2$\pm$10&     0.42$\pm$0.03& - &  4.2   \\
&0.7-0.8&     -21$\pm$28&  1.3$\pm$0.6 &  -0.002$\pm$0.002&      64\\
&0.8-0.9&      -13$\pm$20&      1.39$\pm$0.07& - &   4.3 \\
&0.9-1.&     -36.$\pm$42&  3.$\pm$1.&  -0.004$\pm$0.003&      24\\
        \hline
      
600-700G &0.2-0.3&   0.$\pm$2&    0.057$\pm$0.009& - &    1 \\
      &0.3-0.4&     -2$\pm$3&     0.10$\pm$0.01& - &    1.5 \\
      &0.4-0.5&     -4$\pm$5&     0.15$\pm$0.02& - &     2\\
      &0.5-0.6&     -4.$\pm$5&     0.20$\pm$0.02& - &    1.45\\
      &0.6-0.7&     -3$\pm$7&     0.23$\pm$0.02& - &   4.4\\
      &0.7-0.8&      2$\pm$9&     0.23$\pm$0.03& - & 4.4 \\
      &0.8-0.9&     -8$\pm$13&     0.68$\pm$0.04 & - &  4.2\\
      &0.9-1.&     -6$\pm$17&     0.75$\pm$0.06&&  4.3  \\
      
        \hline
700-800G&0.2-0.3&     0$\pm$1&    0.021$\pm$0.006& - &   2 \\
&0.3-0.4&     2$\pm$2&   -0.03$\pm$0.05 &  0.00029$\pm$  0.00015&      20  \\
&0.4-0.5&    -1$\pm$2&    0.058$\pm$0.008& - &     3.68\\
&0.5-0.6&     -1$\pm$2&    0.087$\pm$0.009& - &     3.1\\
&0.6-0.7&     -1$\pm$3&     0.11$\pm$0.01& - &     3.1\\
&0.7-0.8&     -4$\pm$7&     0.27$\pm$0.15& -0.0005$\pm$0.00045&     20  \\
&0.8-0.9&     -3$\pm$6&     0.28$\pm$0.02& - &    2.8 \\
&0.9-1.&     -3$\pm$9&     0.34$\pm$0.03& - &    2.7 \\
\hline

 
 \end{tabular}}
\end{table}

\section{Dependence on noise threshold} \label{Sec.Noi}
The estimate of geometric properties of magnetic features is dependent on the characteristics of the data 
(such as spatial resolution, scattered-light level, wavelength of observation), 
their calibration and on the details of the analysis performed.
In the context of this study, \citet{chapman2011} showed that the area of faculae and network identified on Ca II K images depends marginally on the 
spatial resolution of the data, and largely on the width of the Ca II K filter used for the observations and the intensity threshold criteria
employed to define magnetic features.  

We therefore repeated previous analysis on data derived employing different magnetic threshold values.
In particular, we compared results obtained with $Th=3\cdot\sigma$ 
with the cases in which $Th=\sigma$, $Th=5\cdot\sigma$, and an arbitrary value of $Th=36G$ constant over the detector.
In general, we found that only the zero order coefficient of the fit, $\alpha$, is somehow affected by the variation of $Th$, while the variations
of the higher order coefficients are typically within the uncertainties of the fits. For instance, when considering 
features located at $0.2<\mu\le1$ and $100G<|B|/\mu\le800G$ we found that $\alpha$ computed on 6-month averaged data varies by $77\%$,  $-50\%$ and $20\%$ 
with respect to results reported in 
Table \ref{tbl-maincoeffs}  
for the cases 
 $Th=\sigma$, $Th=5\cdot\sigma$, $Th=36G$, respectively. 
 
\begin{figure}
 \centerline{\includegraphics[width=6.5cm,trim=6mm 0 0mm 0, clip=]{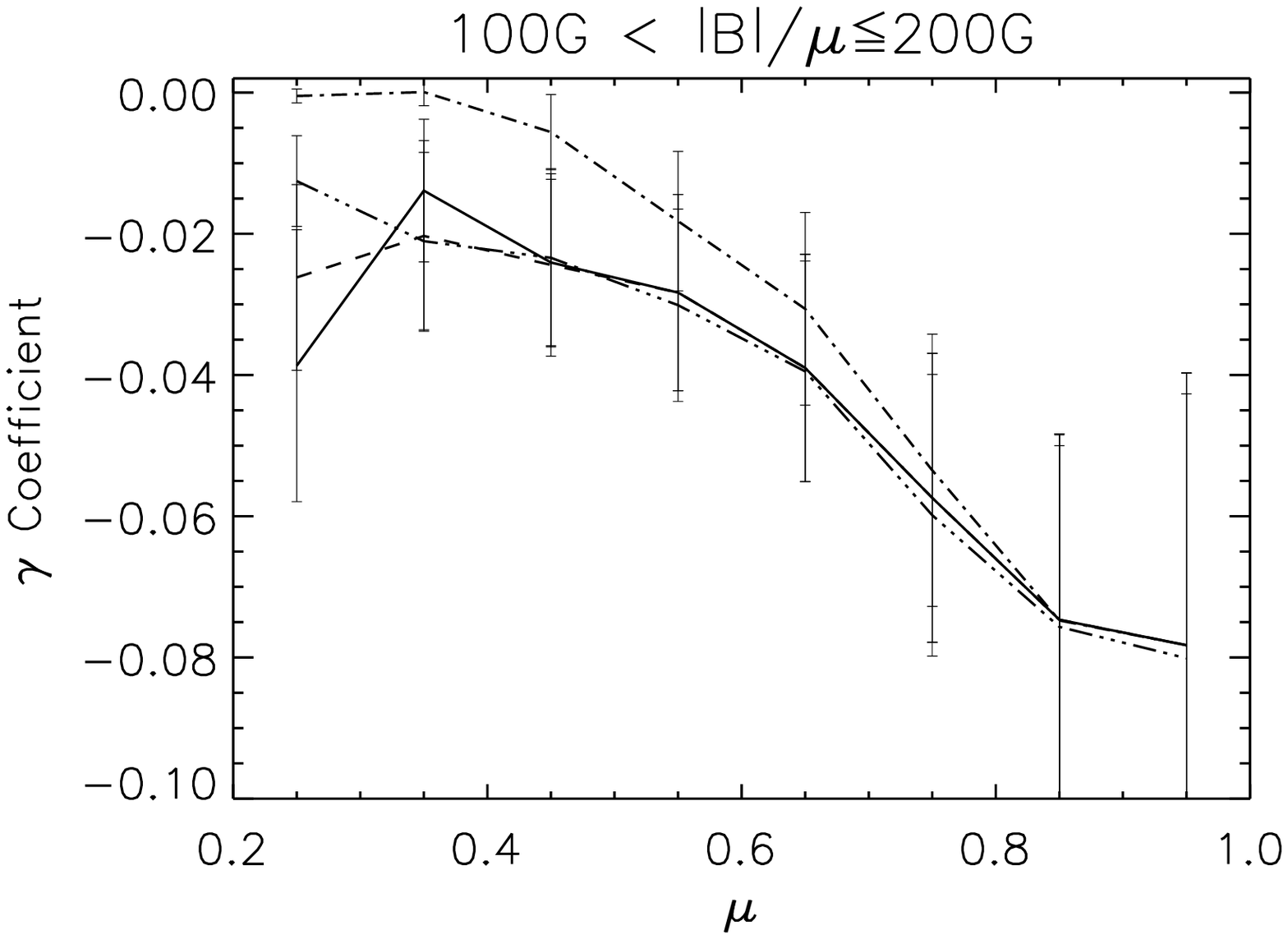}}
 \centerline{\includegraphics[width=6.5cm,trim=10mm 0 0mm 0, clip=]{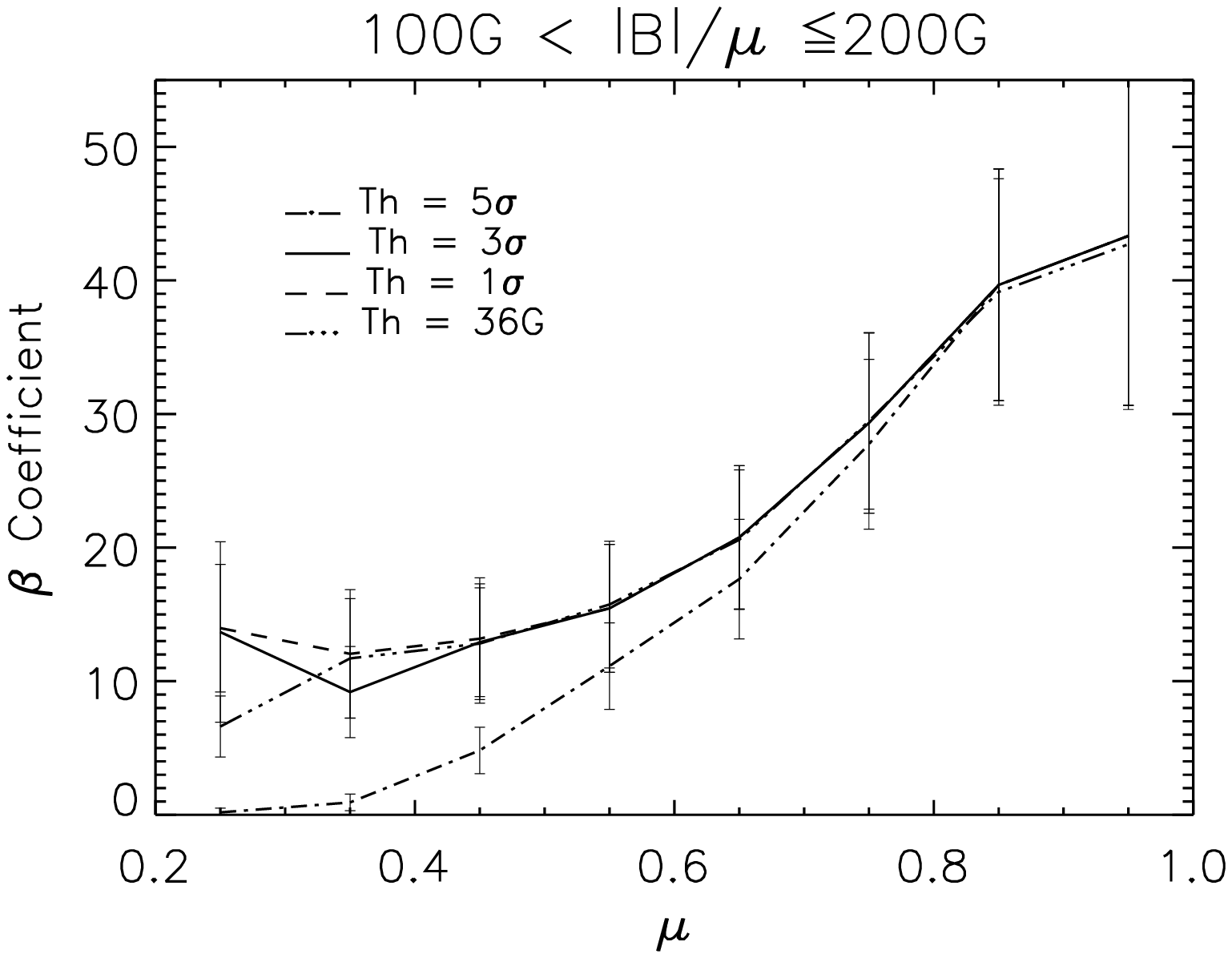}}
\centerline{\includegraphics[width=6.5cm,trim=9mm 0 0mm 0, clip=]{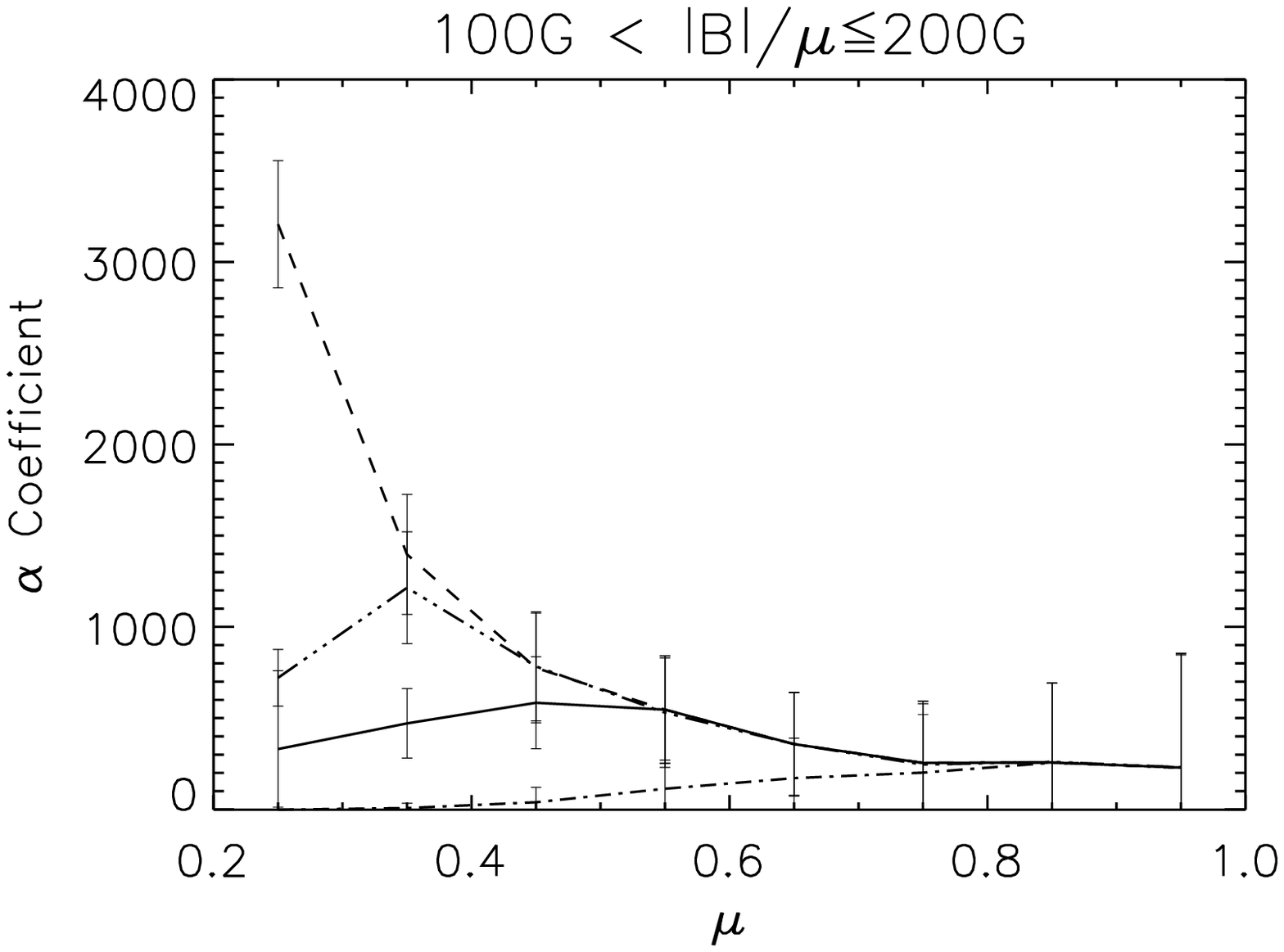}}

\caption{Variation of the quadratic fit coefficients obtained from the daily data with the position on the disk for features with $100G<|B|/\mu\leq 200G$.
Different line-styles represent different noise thresholds. }
\label{fig5}
\end{figure}
 
When restricting the analysis to particular positions on the disk and ranges of magnetic flux values, we found that the
smallest magnetic flux values and the positions closest to the limb are the most affected by variation of the noise threshold value. 
Figure~ \ref{fig5} shows for instance the dependence of
the daily data quadratic fit coefficients on the value of $Th$ for various positions over the disk of features whose magnetic flux satisfies $100G<|B|/\mu\le200G$.
Results obtained for linear fits with the 6-months averaged data are very similar and are not shown.
The plots show that  
all coefficients are sensitive to the noise threshold. The general decrease of both $\beta$ and the absolute value of $\gamma$ toward the limb with the 
increase of $Th$ is the result of 
the change of the statistic within the magnetic flux and $\mu$ bins considered 
and the rapid decrease of $\beta$ and $\gamma$ with the increase of the magnetic flux.
 On the other hand,  the small increase of $\beta$ at $\mu < 0.4$ 
found for $Th = 3 \cdot \sigma$ and $Th = \sigma$  is most likely the result of the rapid decrease of the noise toward the edges of the solar disk.
Moreover, MDI magnetograms acquired before 1998 are known to be slightly defocused, as those data were acquired to favor the focus images acquired in
`high resolution mode''.
 It is therefore likely that the coverage of magnetic elements is slightly underestimated at the minimum, and consequently the slopes of the fits
are slightly overestimated. The effect is expected to be larger toward the limb (because of the compensation of the line-of-sight) and for smaller magnetic 
flux elements (which 
are typically characterized by smaller sizes).

\section{Discussion and Conclusions}\label{Sec.Concl}
In agreement with previous studies carried out on white light and Ca II K images, 
we found that the relation between facular/network area coverage estimated from magnetograms and the sunspot number is linear on data averaged over 6-months, and
is quadratic for daily data. In this last case, the fitted function is characterized by a negative concavity, with the maximum reached toward 
the highest sunspot numbers observed.  

When restricting the analysis to particular ranges of magnetic flux and positions over the disk, the 6-month averaged data are still in general best described by
a linear relation. The daily data are instead best described by a quadratic relation with $\gamma < 0$ for magnetic flux values smaller than 400 G, 
 while for larger magnetic flux values the best fit is returned by a linear function. This is most likely due to the fact that the maxima of the fitted quadratic functions
shift toward higher
sunspot numbers with the increase of the magnetic flux, so that for the larger magnetic flux considered we were able to observe only the rising phase of the relation.

For both 6-months averaged data and daily data, the values of the zero ($\alpha$) and first order ($\beta$) coefficients and the absolute value of the second order coefficient
($\gamma$), decrease with the increase of the magnetic flux. The values of $\beta$ and the absolute value of $\gamma$ increase from the limb toward disk center in all cases.
Instead, the value of $\alpha$ increases from the limb toward disk center when considering linear fits of 6-months averaged data, while no clear dependence on the position over 
the disk is found for fits performed over daily data.

Our results also indicate that for the 6-months averaged data the facular/network and sunspot coverage relation is different during the ascending and the descending phases 
of the cycle.\textbf{ In particular, for a given sunspot number the facular area during the descending phase is larger than during the ascending phase. This is most 
likely consequence of the increase of magnetic field dispersal effects during the descending phase. }
The difference is within
statistical uncertainties when considering all the magnetic features singled out on the magnetograms, but becomes statistically significant when 
restricting the analysis to features located toward disk center, independently on their magnetic flux. Variations of the ratio 
between  facular/network and sunspot areas singled out on Ca II K images were also found during Cycle 23 by \citet{chapman2011}, although those authors
ascribed them to a numerical effects due to the decrease of 
the sunspot area during the periods of minima. Instead  \citet{brownevans1980}, by the analysis of white light faculae,  found differences between the two phases,
 although different cycles showed differences of opposite sign between the ascending and the 
descending phases.
On the contrary, daily data do not show any clear dependence on the phase of the magnetic cycle.

\citet{foukal1998} showed that the different relations between facular/network and sunspot areas obtained on yearly and daily data (binned over sunspot area)
is the result of the different life-times 
of faculae and big sunspots. Observations show indeed that the life-time of faculae increases with the increase of the area of the region at the 
``time of maximum development'',
while the life-time of sunspots shows a saturation at areas larger than 1000 $\mu Hem$ \citep{harvey1993}, so that the decrease of facular area observed
at the largest sunspot areas on the daily data is compensated for in the yearly averaged data. \cite{foukal1998} also pointed out that the daily data describe
best the facular-sunspot area relation close to the emergence of the magnetic flux, which is most likely the reason why the results we obtained on daily data
do not show the differences between
the ascending and descending phases of the cycle that we observed instead for the 6-month averaged data. 

\begin{figure}[h]

\centerline{\includegraphics[scale=.46]{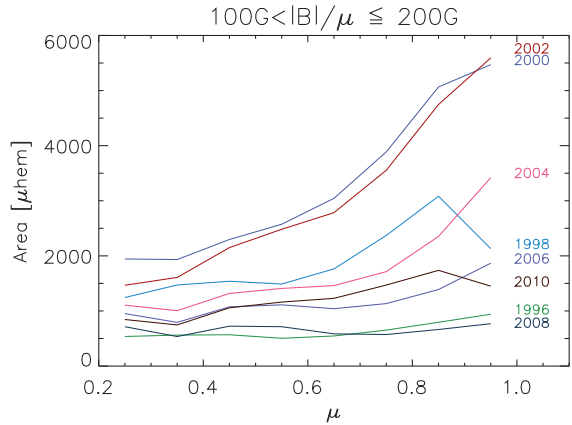}}

\caption{Annual averages of the area coverage at various positions on the disk for features with magnetic flux between 100 and 200G.
For \textbf{simplicity} only data acquired every 
two years are shown.}
\label{fig6}
\end{figure}

In the framework of solar irradiance studies, it is interesting to note that for all magnetic flux values considered the area of magnetic features
located at disk center increases more rapidly than the area of magnetic features located toward the limb. This is illustrated
for instance in Figure~\ref{fig6}, which shows 
the variation with the position $\mu$ of the annual average of features characterized by a magnetic flux value between 100 and 200G. The rapid increase of magnetic 
features area toward disk 
center is due to the larger fraction of the activity belt included in areas close to disk center rather than toward the limb. A small excess is also found toward 
the limb but, although it would be tempting to interpret this result as consequence of the migration of the magnetic field toward the poles, 
we showed in Sec.~\ref{Sec.Noi} that this is most likely an effect of the non-uniformity of the noise toward the edge of the solar disk.

Because the emission of magnetic features
is strongly dependent on the line-of-sight, this result does not necessarily imply that solar irradiance variability is dominated by variations of coverage of features 
located at disk center. In a recent study \citet{criscuoli2014} employed results from magneto hydrodynamic simulations of the solar photosphere to investigate
the contribution of magnetic features to irradiance variations in visible and infrared continua
employing a simple model in which magnetic features are uniformly  distributed over the activity belt. \textbf{We plan to repeat that investigation employing the distributions of
magnetic elements obtained in this study.}

It is important to note that the radiative emission of magnetic features is not only characterized by their magnetic flux, but also by their degree 
of aggregation \citep[e.g.][]{criscuoli2009} and by the properties of the surrounding plasma \citep{solanki1992,kobel2011,romano2012,criscuoli2013}. 
It would be therefore important to repeat the 
present analysis discriminating between pixels located in active or quiet regions and characterized by various degree of aggregation.  

Finally, because previous studies indicate that the area distribution of faculae and network also depends on the strength of the cycle, and to investigate further the 
differences between the ascending and the descending phases, we plan  to extend the present 
analysis to datasets acquired during different cycles. 

\acknowledgments
The author is grateful to Luca Bertello for reading the manuscript and for providing useful comments and suggestions about the analysis of MDI magnetograms.


\end{article}

\end{document}